\documentclass[twocolumn]{revtex4}

\usepackage{ragged2e}
\usepackage{amsmath}
\usepackage{graphicx}

\usepackage[usenames,dvipsnames]{color}

\definecolor{darkblue}{RGB}{0,0,196}

\begin{document}

\title{Rapidity and Energy Dependences of Temperatures and
Volume\\ Extracted from Identified Charged Hadron Spectra\\ in
Proton-Proton Collisions at a Super Proton Synchrotron (SPS)}

\author{Pei-Pin Yang$^{1,}$\footnote{peipinyang@xztu.edu.cn or
peipinyangshanxi@163.com}, Fu-Hu
Liu$^{2,}$\footnote{Correspondence: fuhuliu@163.com;
fuhuliu@sxu.edu.cn}, Khusniddin K.
Olimov$^{3,4,}$\footnote{Correspondence: khkolimov@gmail.com;
kh.olimov@uzsci.net}}

\affiliation{$^1$Department of Physics, Xinzhou Normal University,
Xinzhou 034000, China
\\
$^2$State Key Laboratory of Quantum Optics and Quantum Optics
Devices, Institute of Theoretical Physics, Shanxi University,
Taiyuan 030006, China
\\
$^3$Laboratory of High Energy Physics, Physical-Technical
Institute of Uzbekistan Academy of Sciences, Chingiz Aytmatov Str.
2b, Tashkent 100084, Uzbekistan
\\
$^4$Department of Natural Sciences, National University of Science
and Technology MISIS (NUST MISIS), Almalyk Branch, Almalyk 110105,
Uzbekistan}

\begin{abstract}

\vspace{0.5cm}

\noindent {\bf Abstract:} The standard (Bose-Einstein/Fermi-Dirac
or Maxwell-Boltzmann) distribution from the relativistic ideal gas
model is used to study the transverse momentum ($p_{T}$) spectra
of identified charged hadrons ($\pi^-$, $\pi^+$, $K^-$, $K^+$,
$\bar p$, and $p$) with different rapidities produced in inelastic
proton-proton ($pp$) collisions at the Super Proton Synchrotron
(SPS). The experimental data measured by the NA61/SHINE
Collaboration at the center-of-mass (c.m.) energies
$\sqrt{s}=6.3$, 7.7, 8.8, 12.3, and 17.3 GeV are fitted well by
the distribution. It is shown that the effective temperature
($T_{eff}$ or $T$), kinetic freeze-out temperature ($T_{0}$), and
initial temperature ($T_{i}$) decrease with the increase in
rapidity and increase with the increase in c.m. energy. The
kinetic freeze-out volume ($V$) extracted from the $\pi^-$,
$\pi^+$, $K^-$, $K^+$, and $\bar p$ spectra decreases with the
rapidity and increase with the c.m. energy. The opposite tendency
of $V$, extracted from the $p$ spectra, is observed to be
increasing with the rapidity and decreasing with the c.m. energy
due to the effect of leading protons.
\\

\end{abstract}

\maketitle

\section{Introduction}

The existence of confinement and asymptotic freedom in Quantum
Chromodynamics (QCD) has led to many conjectures about the
thermodynamic and transport properties of the hot and dense
matter. Because of a confinement, a nuclear matter should be
composed of low-energy hadrons, it is considered as a weakly
interacting gas of hadrons. On the other hand, at very high
energies, asymptotic freedom means that the interactions between
quarks and gluons are very weak, and the nuclear matter is
considered as a weakly coupling gas of quarks and gluons. There
should be a phase transition between these two configurations, in
which the degrees of freedom of hadrons disappear and Quark-Gluon
Plasma (QGP) is formed, which is generated at sufficiently high
temperature or density~\cite{1,2,3,4,5,6}. QGP existed in the very
early universe (a few microseconds after the Big Bang), and some
forms of this matter may still exist in the core of neutron stars.
Ultra-relativistic heavy-ion collisions have provided
opportunities to systematically create and study different phases
of the bulk nuclear matter.

Several experiments performed at the Super Proton Synchrotron
(SPS)~\cite{7,8}, Relativistic Heavy Ion Collider
(RHIC)~\cite{2,3,9,10,11,12,13,14,15}, and Large Hadron Collider
(LHC)~\cite{16,17,18,19} have reported abundant experimental data.
The system of proton-proton ($pp$) collisions is usually used as a
reference measurement for heavy ion collisions, as it has several
valence quarks involved in the collisions. Collective flow is one
of the characteristics of the thermal dense medium of this
strongly interacting matter. The generated medium expands
collectively in a way that the flow effect is expected to
distinguish from the thermal motion which reflects the
temperature. The heavy ion physics community has been fascinated
by observing unexpected collective behavior in high multiplicity
$pp$ collision events. It is therefore necessary and important to
study $pp$ collisions.

The transverse momentum ($p_{T}$) spectra of identified charged
hadrons produced in relativistic or high energy collisions contain
abundant information on the collision dynamics and the evolution
properties of the system from the initial stage to the end of
freeze-out phase~\cite{20}. Traditionally, it is believed that
flattening of the $p_{T}$ spectra with high multiplicity is a
signal for the formation of a mixed phase of de-confined partons
and hadrons. In the hydrodynamical model, the slope of $p_{T}$
spectra is co-determined by the kinetic freeze-out temperature and
the transverse expansion flow of the collision system~\cite{21}.
The study of $p_{T}$ spectra can reveal information related to the
effective temperature ($T_{eff}$ or $T$) of the system. A
plateau-like region observed in the excitation function of $T$ is
considered as a possible signal for the formation of mixed-phases,
similar to the temperature dependence of entropy, observed in the
first-order phase transition. In addition, in order to understand
the phase transition from QGP to hadronic matter, the transverse
momentum density is often studied.

In the physical process of high energy heavy ion collisions, at
least four temperatures are often used, namely, initial
temperature ($T_{i}$), chemical freeze-out temperature ($T_{ch}$),
kinetic (or thermal) freeze-out temperature ($T_{0}$), and $T$.
These temperatures correspond to different stages of the
collisions. The excitation degree of the interaction system at the
initial stage is described by $T_i$, at which hadrons undergo
elastic and inelastic interactions in the hadronic medium. Due to
the shortage of the research methods, there is limited research on
$T_i$ in the community, which should be based on the $p_T$. With
the decrease of temperature, the system begins to form hadronic
matter and enters the chemical freeze-out stage. Under the
condition of maintaining a certain degree of local dynamic
equilibrium through quasi elastic resonance scattering, the final
stable hadronic yield has almost no change~\cite{22,23,24,25}. The
$T_{ch}$ and baryon chemical potential ($\mu_B$) at this stage can
be obtained by using various thermodynamic
models~\cite{3,26,27,28}. After chemical freeze-out stage, the
system further expands as the interactions become weak. Finally,
the system enters the kinetic freeze-out stage as the elastic
collisions between hadrons disappear.

In this paper, the $p_{T}$ spectra of identified charged hadrons
($\pi^{-}$, $\pi^{+}$, $K^{-}$, $K^{+}$, $\bar{p}$, and $p$) with
different rapidities produced in inelastic $pp$ collisions at the
center-of-mass (c.m.) energies $\sqrt{s}=6.3$, 7.7, 8.8, 12.3, and
17.3 GeV at the SPS~\cite{29} are studied, where the c.m. energy
is also referred to the collision energy. Although the
nonextensive distribution of the Tsallis
statistics~\cite{37-a,37,38,39,40,41} was widely used in recent
years, the standard (Bose-Einstein/Fermi-Dirac or
Maxwell-Boltzmann) distribution from the relativistic ideal gas
model is still used to extract $T$ directly, and then to obtain
the average transverse momentum ($\langle p_T\rangle$),
root-mean-square transverse momentum ($\sqrt{\langle
p_T^2\rangle}$), $T_0$, and $T_i$, indirectly.

The remainder of this paper is structured as follow. The formalism
and method are described in Section 2. Results and discussion are
provided in Section 3. In Section 4, we summarize our main
observations and conclusions.

\section{Formalism and method}

The particles produced in inelastic $pp$ collisions are thought to
be controlled by two main mechanisms or excitation degrees. The
low-$p_{T}$ region which is less than 1--2 GeV/$c$ is dominated by
the soft excitation process~\cite{30,31}. The high-$p_{T}$ region
which is more than 1--2 GeV/$c$ is governed by the hard scattering
process~\cite{30,31}. The soft process corresponds to a low
excitation degree, and the hard process implies a high excitation
degree. The two-mechanism scheme is only a possible choice in
understanding the particle production. If the particles are
distributed in a very wide $p_T$ region, one should consider the
multiple mechanisms or excitation degrees. If the particles are
distributed in a relative narrow $p_T$ region, one may choose the
single mechanism or excitation degree. In the two-mechanism, it is
currently believed that most light flavor particles are produced
in the soft process. The spectrum in low-$p_{T}$ region shows
exponential behavior, which can be fitted by the thermal
distribution~\cite{32,33,34}. Heavy flavor particles and some
light flavor particles are produced in the hard process. The
spectrum in high-$p_{T}$ region shows inverse power-law behavior
and can be fitted by the Hagedorn~\cite{35,36},
Tsallis--Levy~\cite{37,38}, or Tsallis--Pareto-type
function~\cite{38,39,40,41}.

In this investigation, the light particle spectra in low-$p_T$
region in inelastic $pp$ collisions at the SPS are studied by
using the most basic thermal distribution, the standard
distribution, which comes from the relativistic ideal gas model.
The invariant particle momentum ($p$) distribution described by
the standard distribution can be given by~\cite{37-a}
\begin{align}
E\frac{d^3N}{d^3p} = \frac{1}{2\pi p_T}\frac{d^2N}{dydp_T}=
\frac{gV}{(2\pi)^3}E
\bigg[\exp\bigg(\frac{E-\mu}{T}\bigg)+S\bigg]^{-1},
\end{align}
where $N$ is the particle number, $g$ is the degeneracy factor,
$V$ is the volume, $\mu$ is the chemical potential,
\begin{align}
E=\sqrt{p^2+m_0^2}=m_T\cosh y
\end{align}
is the energy,
\begin{align}
m_T=\sqrt{p_T^2+m_0^2}
\end{align}
is the transverse mass,
\begin{align}
y=\frac{1}{2}\ln\bigg(\frac{1+\beta_z}{1-\beta_z}\bigg)=\tanh^{-1}(\beta_z)
\end{align}
is the rapidity, $\beta_z$ is the longitudinal velocity, and
$S=-1$, $1$, and $0$ correspond to the Bose-Einstein, Fermi-Dirac,
and Maxwell-Boltzmann statistics, respectively.

For the wide $p_T$ spectra, if a multi-component standard
distribution
\begin{align}
E\frac{d^3N}{d^3p} =& \frac{1}{2\pi p_T}\frac{d^2N}{dydp_T}\nonumber\\
=&\sum_{i=1}^{n} \frac{gV_i}{(2\pi)^3}E
\bigg[\exp\bigg(\frac{E-\mu}{T_i}\bigg)+S\bigg]^{-1}
\end{align}
can be used in the fit, one may obtain multiple temperatures, that
is the temperature fluctuation. Here, $n$ denotes the number of
components. Let $k_i$ ($i=1$, 2, ..., $n$) denote the relative
fraction of the $i$-th component, and $V_i$ and $T_i$ are the
volume and temperature corresponding to the $i$-th component
respectively. Naturally, one has
\begin{align}
V=\sum_{i=1}^{n} V_i, \ \ \ T=\sum_{i=1}^{n} k_i T_i, \ \ \
\sum_{i=1}^{n} k_i =1.
\end{align}
Here, $k_i=V_i/V$.

Because of the temperature fluctuation, there are interactions
among different subsystems or local sources due to the exchange of
heat energy. This causes the couplings of entropy functions of
various subsystems. The total entropy is then the sum of the
entropies of subsystems plus the entropies of the couplings. The
temperature fluctuation in the multi-component standard
distribution is a way to explain the origin of Tsallis
distribution. Generally, the $p_T$ spectra which can be fitted by
the multi-component standard distribution can be also fitted by
the Tsallis distribution. Because of the influence of the entropy
index ($q$), the temperature value extracted from the Tsallis
distribution is smaller than that from the multi-component
standard distribution. In fact, in the fit using the Tsallis
distribution, increasing $T$ and/or $q$ can increase the particle
yield in high-$p_T$ region conveniently.

The data sample analyzed in the present work is in the low-$p_T$
region. This implies that the standard distribution can be used.
In the standard distribution, the unit-density function of $y$ and
$p_T$ is written as
\begin{align}
\frac{d^2N}{dydp_T}=& \frac{gV}{(2\pi)^2}p_Tm_T\cosh y \nonumber\\
&\times \bigg[\exp\bigg(\frac{m_T\cosh
y-\mu}{T}\bigg)+S\bigg]^{-1}.
\end{align}
Then, the density function of $p_T$ is
\begin{align}
\frac{dN}{dp_T}=& \frac{gV}{(2\pi)^2}p_Tm_T\int_{y_{\min}}^{y_{\max}}\cosh y \nonumber\\
&\times \bigg[\exp\bigg(\frac{m_T\cosh
y-\mu}{T}\bigg)+S\bigg]^{-1} dy,
\end{align}
where $y_{\min}$ and $y_{\max}$ are the minimum and maximum
rapidities in the rapidity interval, respectively. The density
function of $y$ is
\begin{align}
\frac{dN}{dy}=& \frac{gV}{(2\pi)^2}\cosh y \int_{0}^{p_{T\max}}p_Tm_T \nonumber\\
&\times \bigg[\exp\bigg(\frac{m_T\cosh
y-\mu}{T}\bigg)+S\bigg]^{-1} dp_T,
\end{align}
where $p_{T\max}$ is the maximum $p_T$ in the considered rapidity
interval. Although $p_{T\max}$ can be mathematically an infinity,
it is only large enough in physics due to the limitations from the
energy and momentum conservation.

No matter what the specific form of particle momentum distribution
is used, the probability density function of $p_T$ is written in
general as
\begin{align}
f(p_T)=&\frac{1}{N}\frac{dN}{dp_T}.
\end{align}
Naturally, $f(p_T)$ is normalized to 1. That is
\begin{align}
\int_0^{\infty}f(p_T)dp_T=1.
\end{align}
One has the average transverse momentum,
\begin{align}
\langle p_{T}\rangle
=\frac{\int_{0}^{\infty}p_{T}f(p_{T})dp_{T}}{\int_{0}^{\infty}f(p_{T})dp_{T}}
=\int_{0}^{\infty}p_{T}f(p_{T})dp_{T},
\end{align}
and the root-mean-square $p_T$,
\begin{align}
\sqrt{\langle p^2_T\rangle}
=\sqrt{\frac{\int_{0}^{\infty}p^2_{T}f(p_{T})dp_{T}}{\int_{0}^{\infty}f(p_{T})dp_{T}}}
=\sqrt{\int_{0}^{\infty}p^2_{T}f(p_{T})dp_{T}}.
\end{align}

In principle, there are three independent chemical potentials:
baryon ($\mu_B$), electric charge or isospin ($\mu_I$), and
strangeness ($\mu_S$) which are related to the three conserved
charges. Although the chemical potential, $\mu_{\pi}$ ($\mu_{K}$
or $\mu_p$), of the pion (kaon or proton) can be written in terms
of the above three chemical
potentials~\cite{42,43,44,45,46,47,48}, they are obtained by us
using an alternative method in the present work for more
convenience.

Considering the yield ratio [$k_j$ ($j=\pi$, $K$, and $p$)] of
negative to positive charged hadrons ($j^-$ to $j^+$), the
corresponding chemical potentials ($\mu_{j^-}$ and $\mu_{j^+}$),
as well as the corresponding source temperature ($T_{j^-}$ and
$T_{j^+}$), one has the relationship between $k_j$ and $\mu_{j}$
to be~\cite{20,48a,48b,48c,48d}
\begin{align}
k_j\equiv\frac{j^-}{j^+}=\exp\left(\frac{\mu_{j^-}}{T_{j^-}}-\frac{\mu_{j^+}}{T_{j^+}}\right)
=\exp\left(-\frac{2\mu_j}{T_j}\right)
\end{align}
if the condition
\begin{align}
T_{j^-}=T_{j^+}=T_j, \ \ \ \mu_{j^-}=-\mu_{j^+}=-\mu_{j}
\end{align}
are satisfied. Here, $j^-$ and $j^+$ also denote the yields of
negative and positive hadrons respectively. $k_j$ can be obtained
simply from the experimental data, $T_j$ should be the chemical
kinetic-freeze temperature $T_{ch}$ which is slightly larger than
or equal to the effective temperature $T$ due to the short
lifetime of the system formed in $pp$ collisions. One has
$T_j\approx T$ in this work.

Further, one has
\begin{align}
\mu_j=-\frac{1}{2}T_j\ln k_j.
\end{align}
Obviously, $\mu_j$ is energy dependent due to $T_j$ and $k_j$
being energy dependent. Based on a collection of large amounts of
experimental data, our previous work~\cite{48c,48d} presents the
excitation functions of $\mu_{j}$ in $pp$ and central heavy ion
collisions, which can be used for a direct extraction for this
study. In particular, $\mu_j$ decreases quickly with the increase
of energy in $pp$ collisions in the concerned SPS energy range.
However, the tendency of $\mu_{\pi}$ in central heavy ion
collisions is opposite to that in $pp$ collisions, though the
tendency of $\mu_K$ is similar and that of $\mu_p$ is also similar
in the two collisions. The three $\mu_j$ in both the collisions
are close to 0 at around 100 GeV and above.

The chemical freeze-out temperature $T_{ch}$ in central heavy ion
collisions is also energy dependent~\cite{42,43,44,45,46,47,48},
which shows a tendency of a rapid increase at a few GeV, and then
saturation at dozens of GeV and above. In view of the fact that
the tendency of $T_{ch}$ has a parameterized excitation function
with unanimity in the community, the present work does not study
$T_{ch}$ parameter.

Generally, the kinetic freeze-out temperature $T_0$ has a tendency
of a rapid increase at a few GeV and then an ambiguous tendency
(increase, decrease, or saturation) appears at dozens of GeV and
above. It is worth studying the tendency of $T_0$ further. A
thermal-related method shows that~\cite{49}
\begin{align}
T_{0}=\frac{\langle p_{T}\rangle}{2\kappa_{0}},
\end{align}
where $\kappa_{0}=3.07$ is a coefficient and 2 is introduced by us
because two participant partons (one from the projectile and the
other from the target) are assumed to contribute to $\langle
p_{T}\rangle$. This formula gives an approximate consistent
tendency of $T_0$ as another thermal-related method~\cite{50}
which shows $T_0$ to be proportional to $\langle p_T\rangle$ and
the coefficient to be energy-related, though the results from the
two methods are not the same.

\begin{figure*}[htbp]
\vskip-0.5cm \hspace{0.cm}
\includegraphics[width=17.25cm]{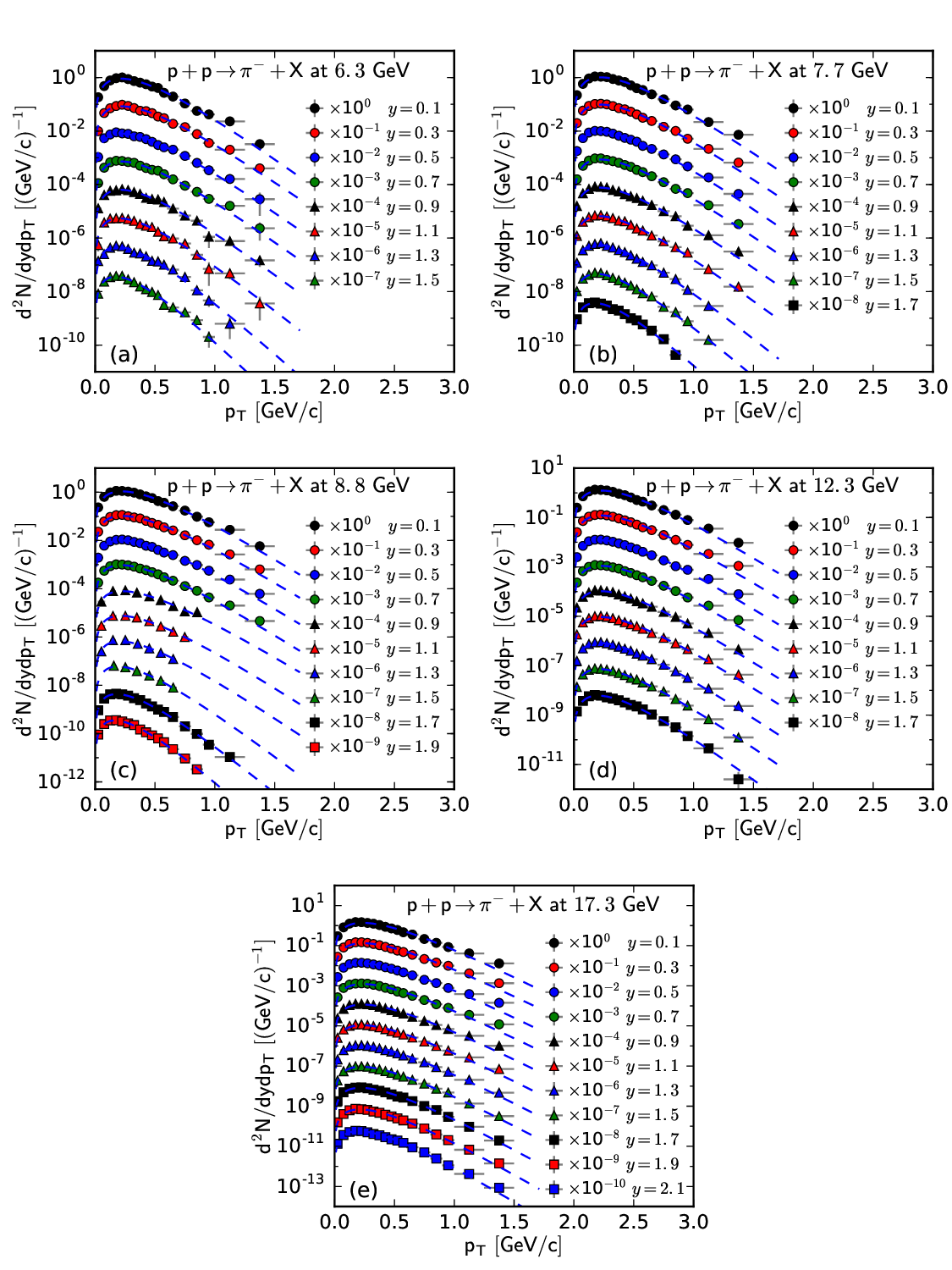}
\vskip.2cm \justifying\noindent {Fig 1. The spectra of $\pi^{-}$
produced in $pp$ collisions at $\sqrt{s}=$ (a) 6.3, (b) 7.7, (c)
8.8, (d) 12.3, and (e) 17.3 GeV at different $y$ with an interval
width of 0.2. The symbols represent the experimental data measured
by the NA61/SHINE Collaboration~\cite{29} and the curves are the
fitting results by the Bose-Einstein distribution.}
\end{figure*}

\begin{figure*}[htbp]
\vskip-1.5cm \hspace{0.cm}
\includegraphics[width=17.25cm]{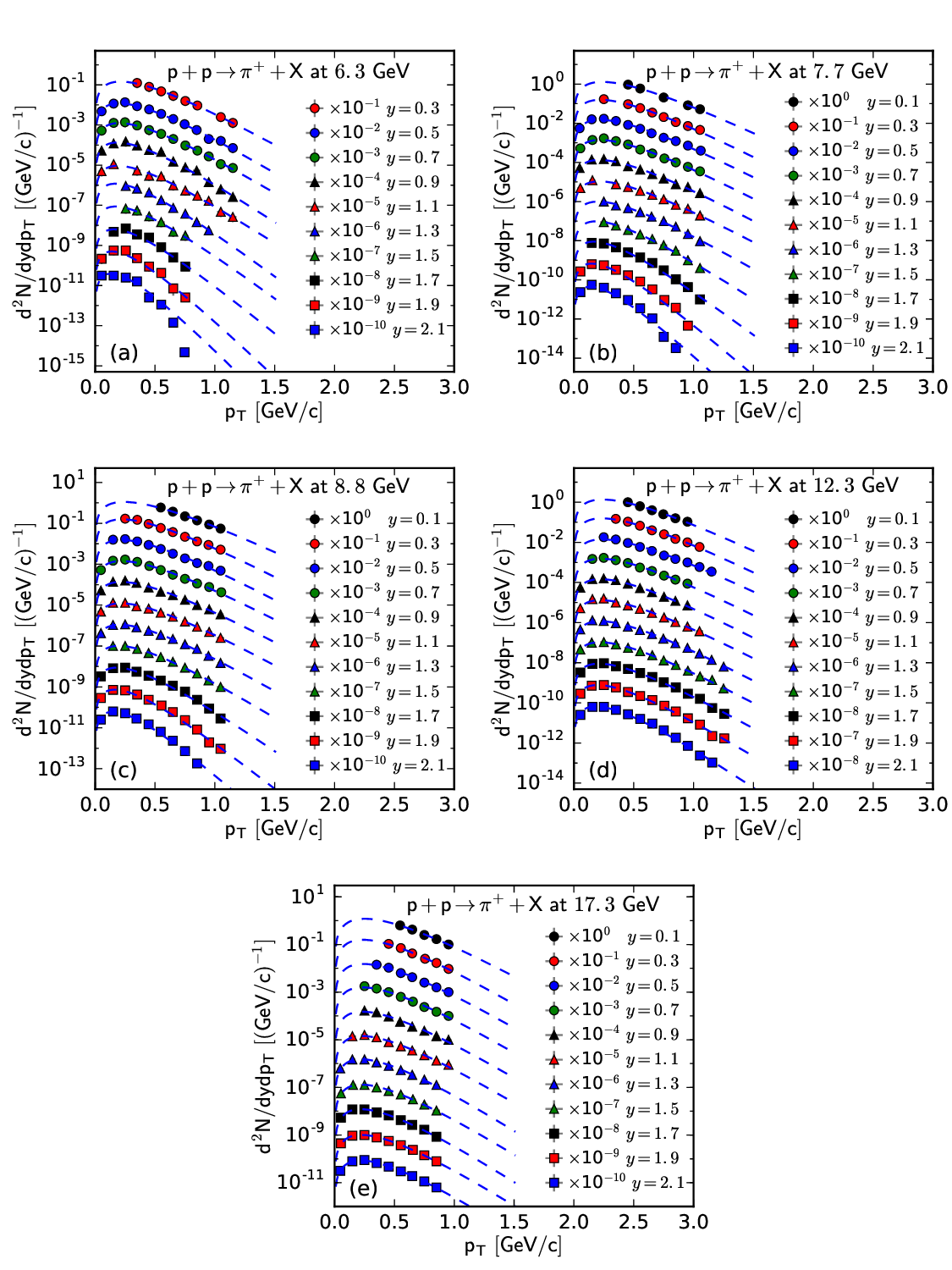}
\vskip0.2cm \justifying\noindent {Fig. 2. Same as Figure 1, but
for $\pi^+$ production.}
\end{figure*}

\begin{figure*}[htbp]
\vskip-1.5cm \hspace{0.cm}
\includegraphics[width=17.25cm]{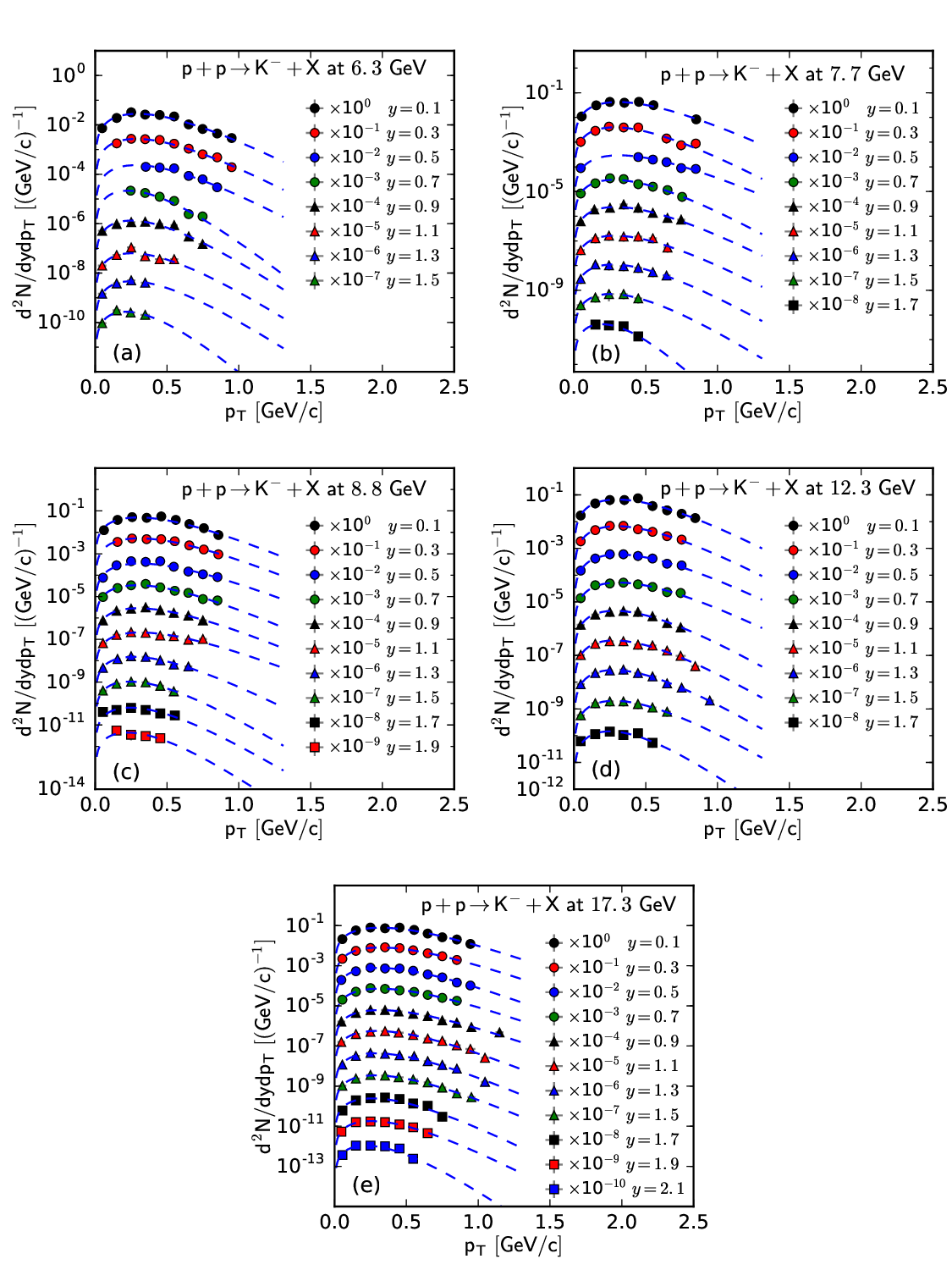}
\vskip0.2cm \justifying\noindent {Fig. 3. Same as Figure 1, but
for $K^-$ production.}
\end{figure*}

\begin{figure*}[htbp]
\vskip-1.5cm \hspace{0.cm}
\includegraphics[width=17.25cm]{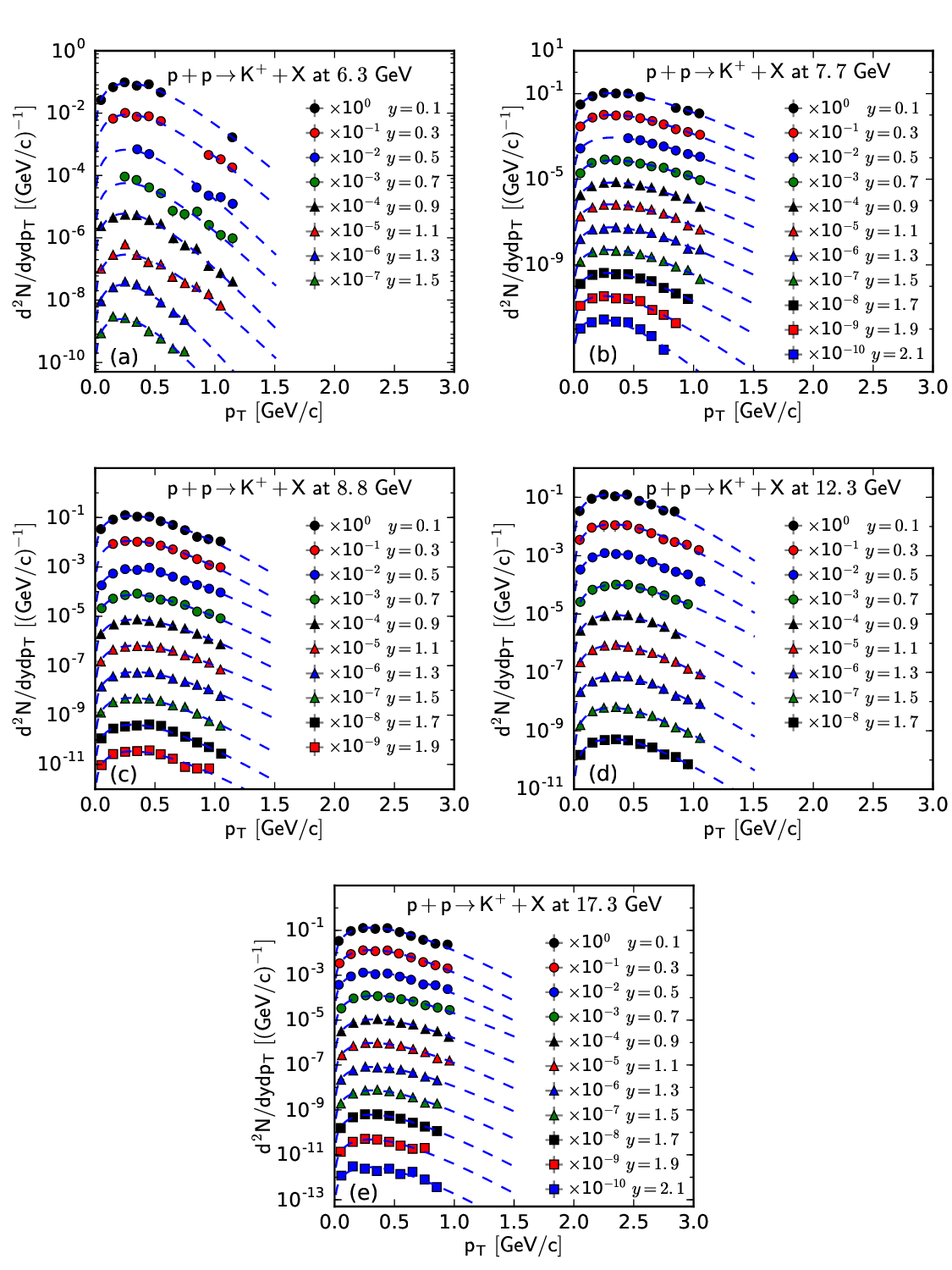}
\vskip0.2cm \justifying\noindent {Fig. 4. Same as Figure 1, but
for $K^+$ production.}
\end{figure*}

\begin{figure*}[htb!]
\vskip-0.35cm \hspace{0.cm}
\includegraphics[width=17.25cm]{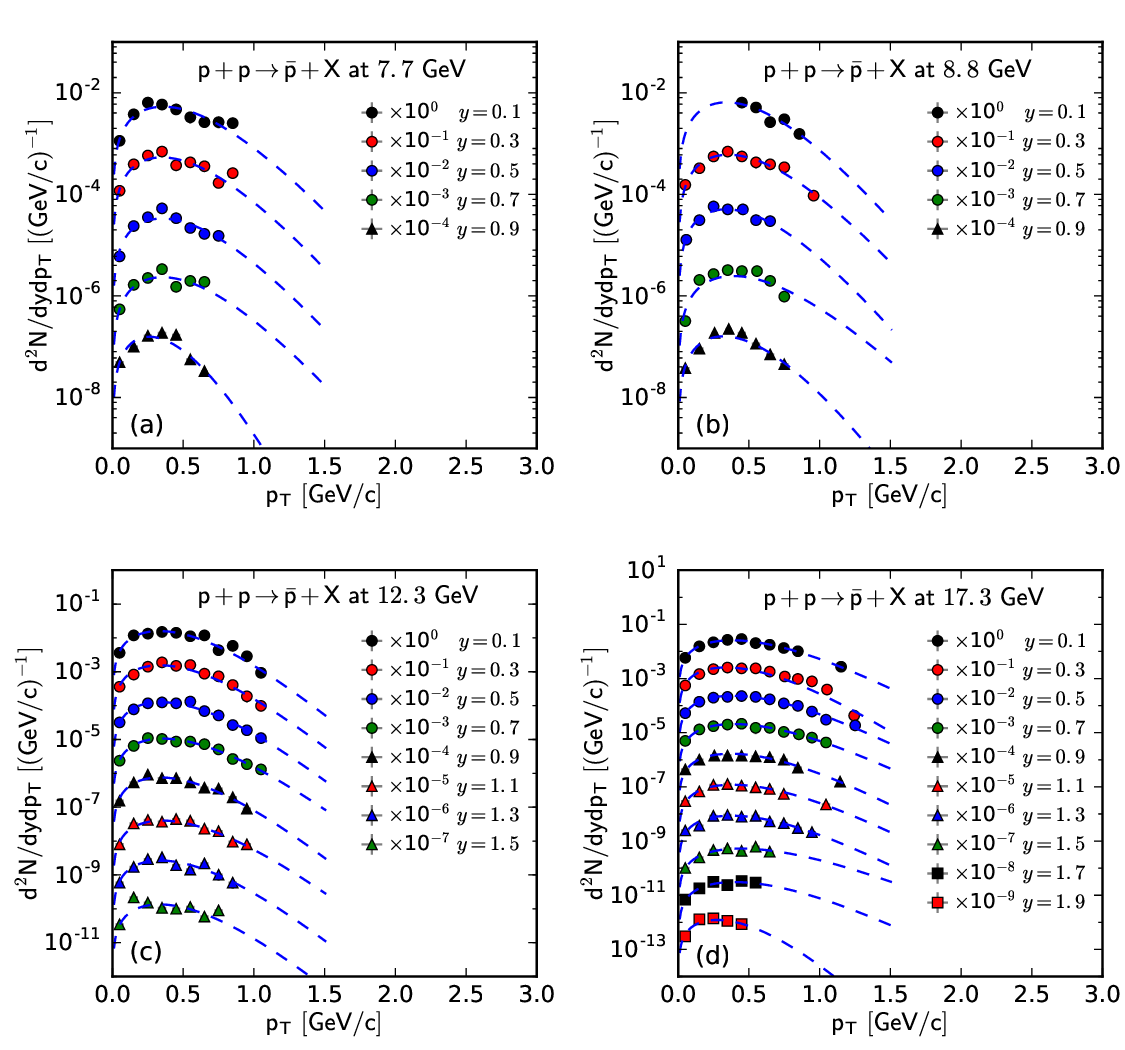}
\vskip0.2cm \justifying\noindent {Fig. 5. Same as Figure 1, but
for $\bar p$ production, where the data at $\sqrt{s}=6.3$ GeV is
not available in experiments. The curves are the fitting results
by the Fermi-Dirac distribution.}
\end{figure*}

\begin{figure*}[htbp]
\vskip-1.5cm \hspace{0.cm}
\includegraphics[width=17.25cm]{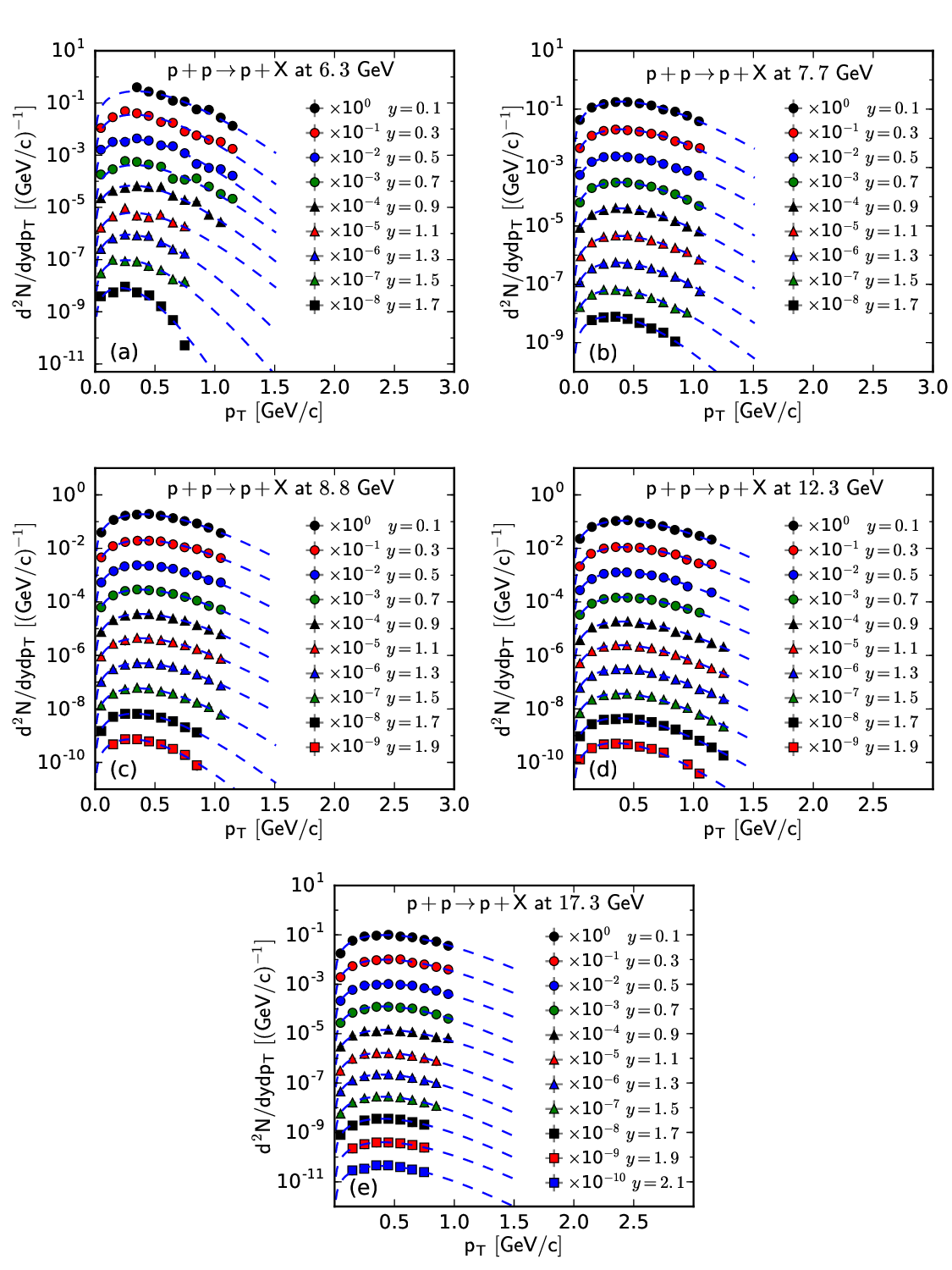}
\vskip0.2cm \justifying\noindent {Fig. 6. Same as Figures 1 and 5,
but for $p$ production.}
\end{figure*}

The initial temperature $T_i$, which is comparable to the
experimental data, is less studied in the community. According to
the string percolation model~\cite{51,52,53}, $T_i$ is expressed
as
\begin{align}
T_{i}=\sqrt{\frac{\langle p_{T}^{2}\rangle}{2F(\xi)}},
\end{align}
where
\begin{align}
F(\xi)=\sqrt{\frac{1-\exp(-\xi)}{\xi}}
\end{align}
is the color suppression factor related to the dimensionless
percolation density parameter $\xi$. In $pp$ collisions,
$F(\xi)\sim 1$ due to the low string overlap probability. As an
initial quantity, $T_i$ should reflect the excitation degree of
the system at the parton level. Correspondingly, the final
quantity $T_0$ should also be extracted at the parton level. This
is also the reason that 2 is introduced by us in the denominator
of $T_0$ expression if one assumes that two participant partons
are the energy sources in the formation of a particle.

The kinetic energy of a particle directional movement should not
be reflected in the temperature parameters. The experimental data
used in this paper were all measured in the forward rapidity
region. In order to remove the influence of directional motion,
one can directly shift the forward rapidity and its interval to
the mid-rapidity with the same interval width during the fitting
process. In this paper, we integrate $y$ from $y_{\min}=-0.1$ to
$y_{\max}=0.1$ in the fit to give more accurate result, though
$y\approx0$ and $\cosh y\approx 1$ near the mid-rapidity. The
small difference ($<1\%$) between the accurate and approximate
calculations appears mainly in the normalization, but not the
temperature parameter.

The method of least squares based on obtaining the minimum
$\chi^2$ is adopted to obtain the best parameters and their
uncertainties. The treatment method is given in appendix A.

\section{Results and discussion}

Figures 1 and 2 show the rapidity dependent double differential
$p_{T}$ spectra, $d^2N/dydp_{T}$, of $\pi^{-}$ and $\pi^{+}$
respectively, produced in inelastic $pp$ collisions at the SPS.
Panels (a)--(e) correspond to the results of $\sqrt{s}=6.3$, 7.7,
8.8, 12.3, and 17.3 GeV, respectively. The symbols represent the
experimental data at different $y$, with an interval width of 0.2
unit, measured by the NA61/SHINE Collaboration~\cite{29}, and the
curves are our results fitted from the Bose-Einstein distribution.
In order to see the fitting effect more clearly, the experimental
data and fitting results at different rapidities are multiplied by
different factors labeled in the panel for scaling. The values of
related free parameters ($T$), normalization constant ($V$),
$\chi^2$, and number of degrees of freedom (ndof) for the curves
in Figures 1 and 2 are listed in Table 1 in appendix B. One can
see that the fitting results with the Bose-Einstein distribution
are in good agreement with the experimental data of $\pi^{-}$ and
$\pi^{+}$ spectra, measured by the NA61/SHINE Collaboration in
$pp$ collisions at different $\sqrt{s}$ and in different $y$
intervals.

Similar to Figures 1 and 2, Figures 3 and 4 show the rapidity
dependent $d^2N/dydp_{T}$ of $K^{-}$ and $K^{+}$, respectively,
produced in inelastic $pp$ collisions at different $\sqrt{s}$. The
values of $T$, $V$, and $\chi^2$/ndof for the curves in Figures 3
and 4 are listed in Table 2 in appendix B. One can see that the
fitting results by the Bose-Einstein distribution are in agreement
with the experimental data of $K^{-}$ and $K^{+}$, measured by the
NA61/SHINE Collaboration in $pp$ collisions at different
$\sqrt{s}$ and in different $y$ intervals.

Similar to Figures 1--4, Figures 5 and 6 show the rapidity
dependent $d^2N/dydp_{T}$ of $\bar{p}$ and $p$ respectively,
produced in inelastic $pp$ collisions at different $\sqrt{s}$. The
experimental data of $\bar{p}$ at $\sqrt{s}=6.3$ GeV in Figure 5
is not available. The values of $T$, $V$, and $\chi^2$/ndof for
the curves in Figures 5 and 6 are listed in Table 3 in appendix B.
One can see that the $p_{T}$ spectra of $\bar{p}$ and $p$ in $pp$
collisions are shown to obey approximately the Fermi-Dirac
distribution.

To show more intuitively the dependence of the free parameter $T$
and derived quantities (the kinetic freeze-out temperature $T_{0}$
and initial temperature $T_{i}$) on rapidity, $y$, and c.m.
energy, $\sqrt{s}$, Figures 7--10 show the relations of $T$--$y$,
$T_0$--$y$, $T_i$--$y$, and $V$--$y$ at different $\sqrt{s}$,
respectively, and Figures 11--14 show the relations of
$T$--$\sqrt{s}$, $T_0$--$\sqrt{s}$, $T_i$--$\sqrt{s}$, and
$V$--$\sqrt{s}$ at different $y$, respectively. Panels (a)--(f)
correspond to the results from $\pi^{-}$, $\pi^{+}$, $K^{-}$,
$K^{+}$, $\bar{p}$, and $p$ spectra, respectively. These figures
show some changing trends of parameters.

In most cases, one can generally see that $T$, $T_0$, and $T_i$
decrease (increase) with the increase in $y$ ($\sqrt{s}$). There
is a tendency of saturation for the three temperatures at
$\sqrt{s}=7.7$ GeV and above. Being the initial energy of a
saturation effect, 7.7 GeV is a special energy at which the
reaction products are proton-dominated and above which the
products are meson-dominated. For $\pi^{-}$, $\pi^{+}$, $K^{-}$,
$K^{+}$, and $\bar{p}$ spectra, the extracted $V$ also decreases
(increases) with the increase in $y$ ($\sqrt{s}$). However, for
$p$ spectra, the extracted $V$ shows an opposite tendency,
increasing (decreasing) with the increase in $y$ ($\sqrt{s}$).

\begin{figure*}[htbp]
\vskip-1.5cm \hspace{0.cm}
\includegraphics[width=17.25cm]{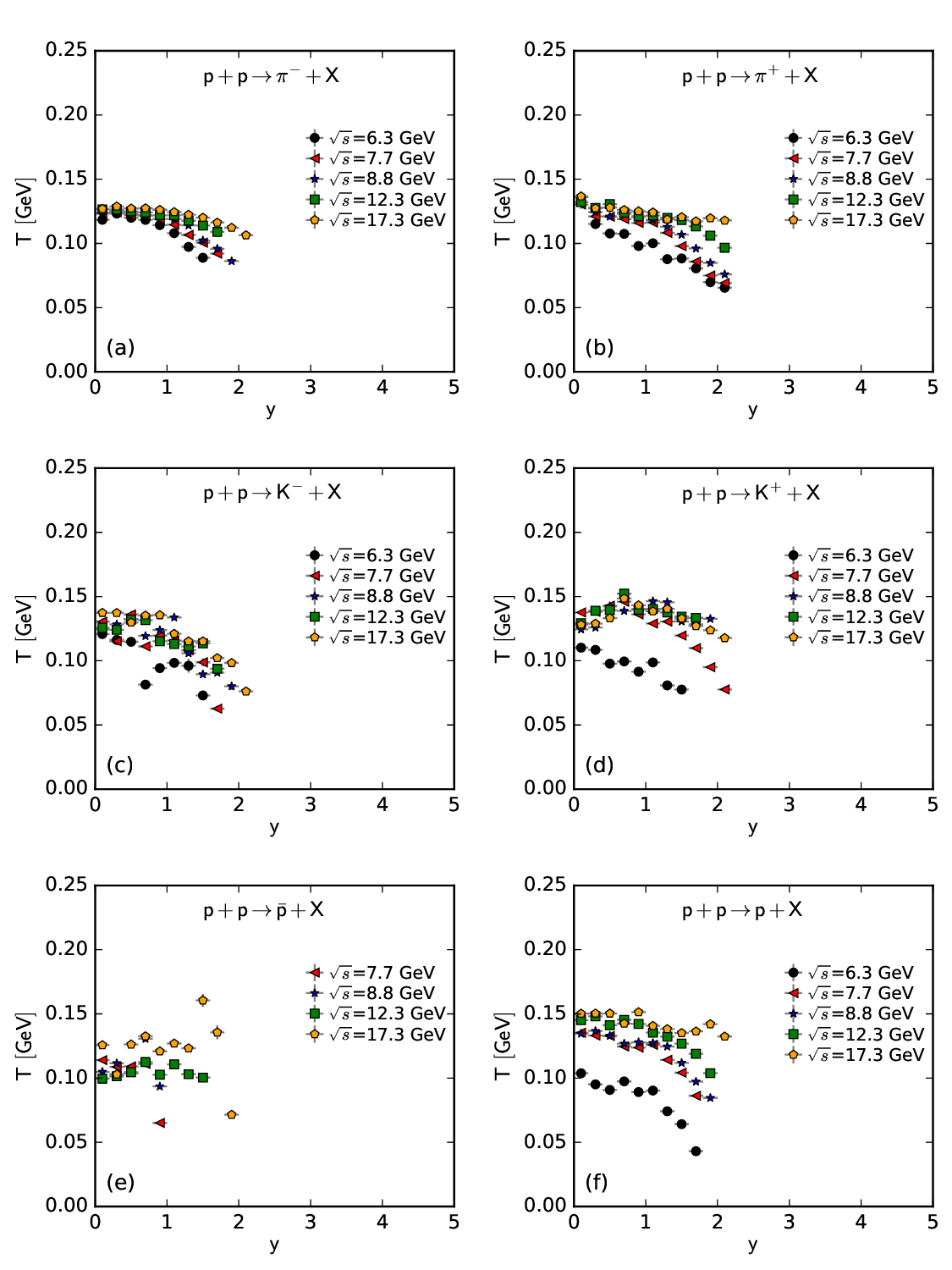}
\vskip0.2cm \justifying\noindent {Fig. 7. Dependence of $T$ on $y$
at different $\sqrt{s}$ from the spectra of (a) $\pi^{-}$, (b)
$\pi^{+}$, (c) $K^{-}$, (d) $K^{+}$, (e) $\bar{p}$, and (f) $p$.}
\end{figure*}

\begin{figure*}[htbp]
\vskip-1.5cm \hspace{0.cm}
\includegraphics[width=17.25cm]{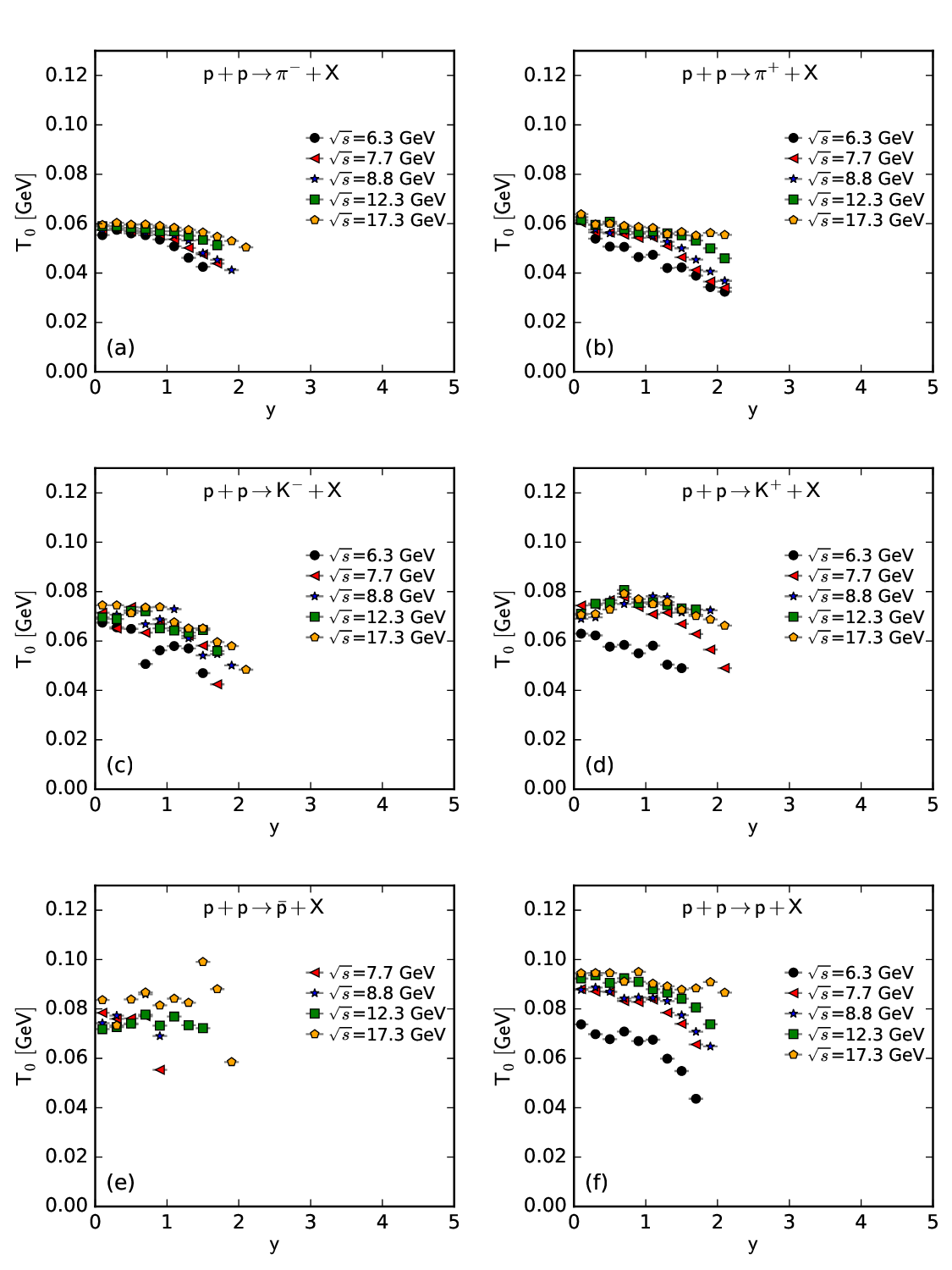}
\vskip0.2cm \justifying\noindent {Fig. 8. Same as Figure 7, but
for the dependence of $T_0$ on $y$ at different $\sqrt{s}$.}
\end{figure*}

\begin{figure*}[htbp]
\vskip-1.5cm \hspace{0.cm}
\includegraphics[width=17.25cm]{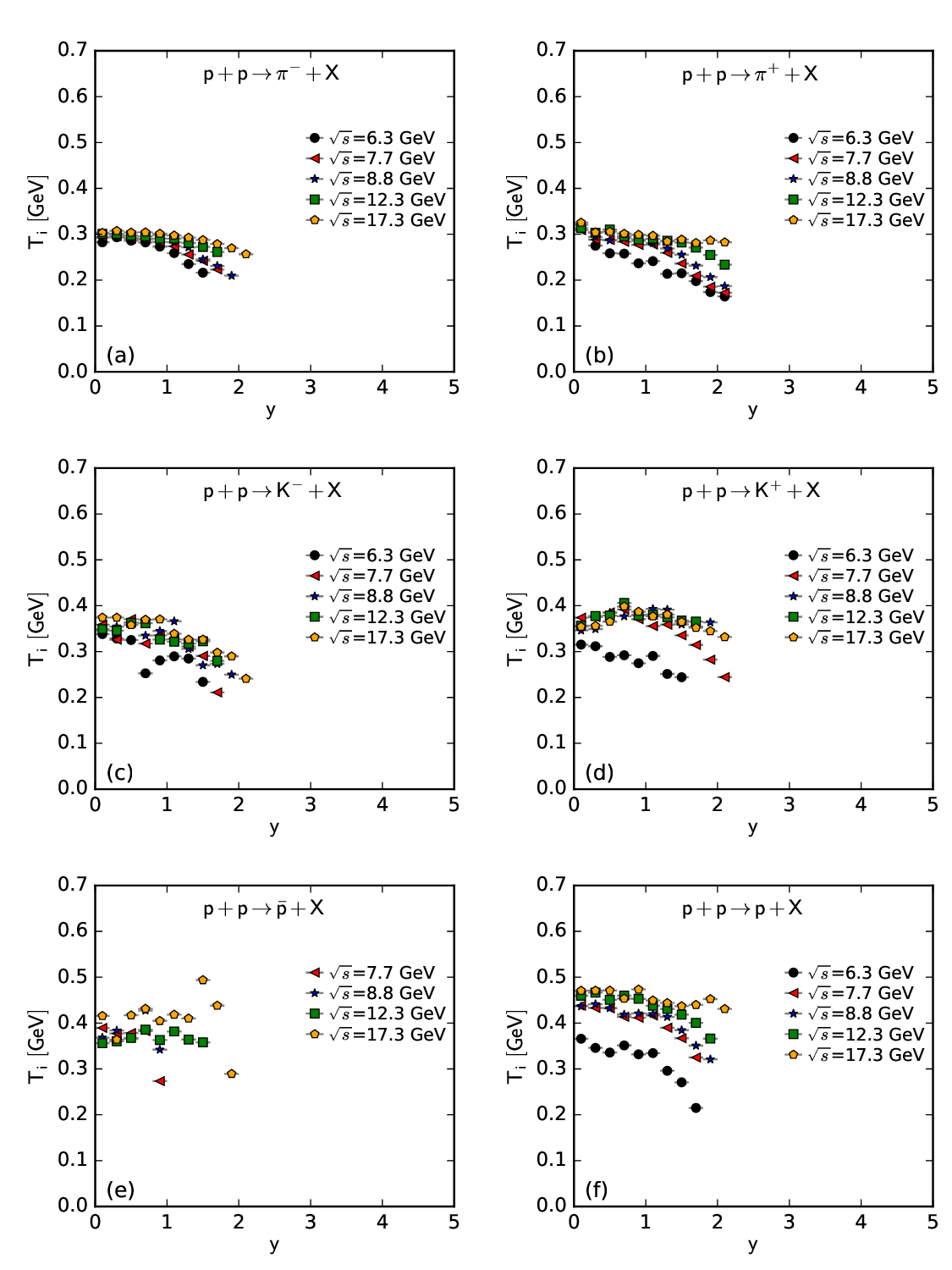}
\vskip0.2cm \justifying\noindent {Fig. 9. Same as Figure 7, but
for the dependence of $T_i$ on $y$ at different $\sqrt{s}$.}
\end{figure*}

\begin{figure*}[htbp]
\vskip-1.5cm \hspace{0.cm}
\includegraphics[width=17.25cm]{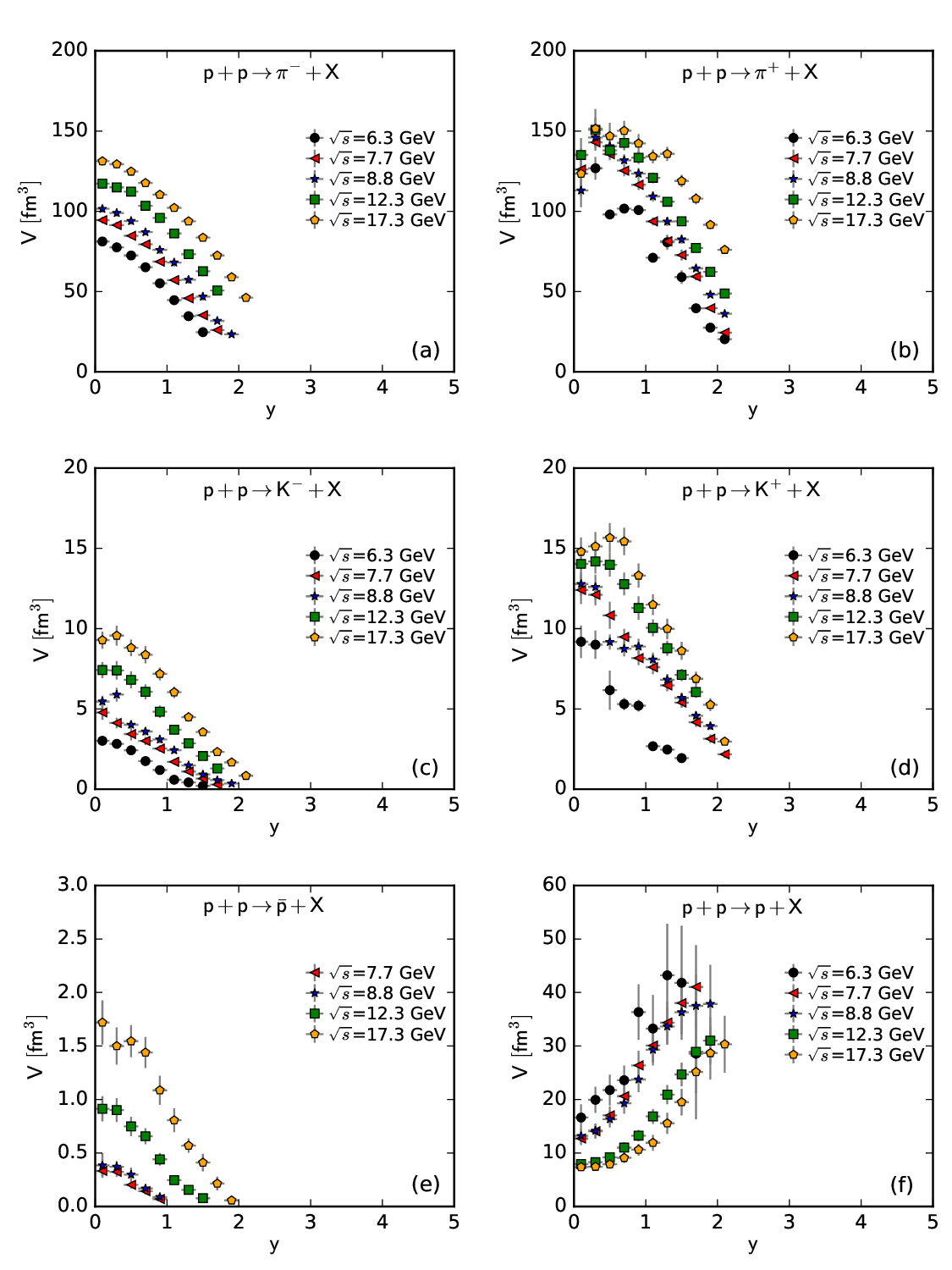}
\vskip0.2cm \justifying\noindent {Fig. 10. Same as Figure 7, but
for the dependence of $V$ on $y$ at different $\sqrt{s}$.}
\end{figure*}

There is an isospin and mass independence of $T$. This property is
exactly that of $T_{ch}$, which implies a single scenario of
chemical freeze-out. However, although $T_0$ and $T_i$ are isospin
independent, they increase with the increase in mass. The
mass-dependence of $T_0$ is a reflection of mass-dependent
differential kinetic freeze-out scenario or multiple kinetic
freeze-out scenario. The mass-dependence of $T_i$ means that the
formation moments of different particles are different. With the
increase of $T_0$ ($T_i$), massive particles are emitted (formed)
earlier. Averagely, this work shows that $\bar p(p)$ are emitted
(formed) earlier than $K^{\mp}$, and $K^{\mp}$ are emitted
(formed) earlier than $\pi^{\mp}$, though the relaxation times for
the emissions (formations) of different particles can overlap.

Except for $V$ from the $p$ spectra, the tendencies of other
parameters from the $p$ spectra, and the tendencies of parameters
from the spectra of other particles are easy to understand. It is
expected that the local system in the mid-rapidity region has more
deposited energy than that in the forward region. Meanwhile, the
collision system at higher energy has more deposited energy than
that at lower energy. This results in a higher excitation degree
(then higher temperature) at the mid-rapidity and more produced
particles (then larger volume) at higher energy.

The $V$ tendency from the $p$ spectra is opposite to that from the
spectra of other particles. The reason is that the pre-existing
leading protons affect the $p$ spectra. Because of the leading
protons appearing in the forward region, the number of protons and
then the volume of a proton source in the fixed interval are small
at the mid-rapidity. At higher energy, the leading protons appear
in the more forward region, which leads to a smaller $V$ in the
fixed interval in the rapidity space. In the present work, the
fixed interval is that $\Delta y=y_{\max}-y_{\min}=0.2$.

The values of $V$ depend on particle mass and charge. Excluding
the case of $p$, which contains pre-existing leading protons in
$pp$ system, $V$ decreases significantly with the increase in
mass, and positive hadrons correspond to the larger $V$ of
emission source. This is because the larger the mass, the more
difficult to produce this particle. Meanwhile, there is an
electromagnetic exclusion (attraction) between positive (negative)
hadrons and pre-existing protons. This causes larger (smaller) $V$
of emission source of positive (negative) hadrons.

Generally, the effective temperature $T$ is proportional to the
mean transverse momentum $\langle p_{T}\rangle$. The present work
shows that $T_{\pi^{-}}\approx0.351 \langle
p_{T}\rangle_{\pi^{-}}$, $T_{\pi^{+}}\approx0.348 \langle
p_{T}\rangle_{\pi^{+}}$, $T_{K^{-}}\approx0.284 \langle
p_{T}\rangle_{K^{-}}$, $T_{K^{+}}\approx0.293 \langle
p_{T}\rangle_{K^{+}}$, $T_{\bar{p}}\approx0.234 \langle
p_{T}\rangle_{\bar{p}}$, and $T_{p}\approx0.240 \langle
p_{T}\rangle_p$. Here, the type of a particle appears as the
subscript label of the related quantity. The ratio of $T/\langle
p_{T}\rangle$ is approximately independent of a particle mass.
This is consistent to the ratios of $T_0/\langle p_{T}\rangle$ and
$T_i/\sqrt{\langle p_{T}^{2}\rangle}$, which are independent of
particle mass according to Eqs. (17) and (18).

As only a free parameter, $T$ does not show an obvious dependence
on particle type or mass. However, it is hard to extract exact
information from $T$ because it is not a real temperature, because
it also contains the contribution of transverse flow. $T_0$ is
smaller than $T_i$ due to the fact that $T_0$ is ``measured" at
the kinetic freeze-out stage (the final one), and $T_i$ is
``measured" at the initial stage. From the initial stage to the
final one, the system becomes colder and colder. This is indeed
observed in the present work.

\begin{figure*}[htbp]
\vskip-1.5cm \hspace{0.cm}
\includegraphics[width=17.25cm]{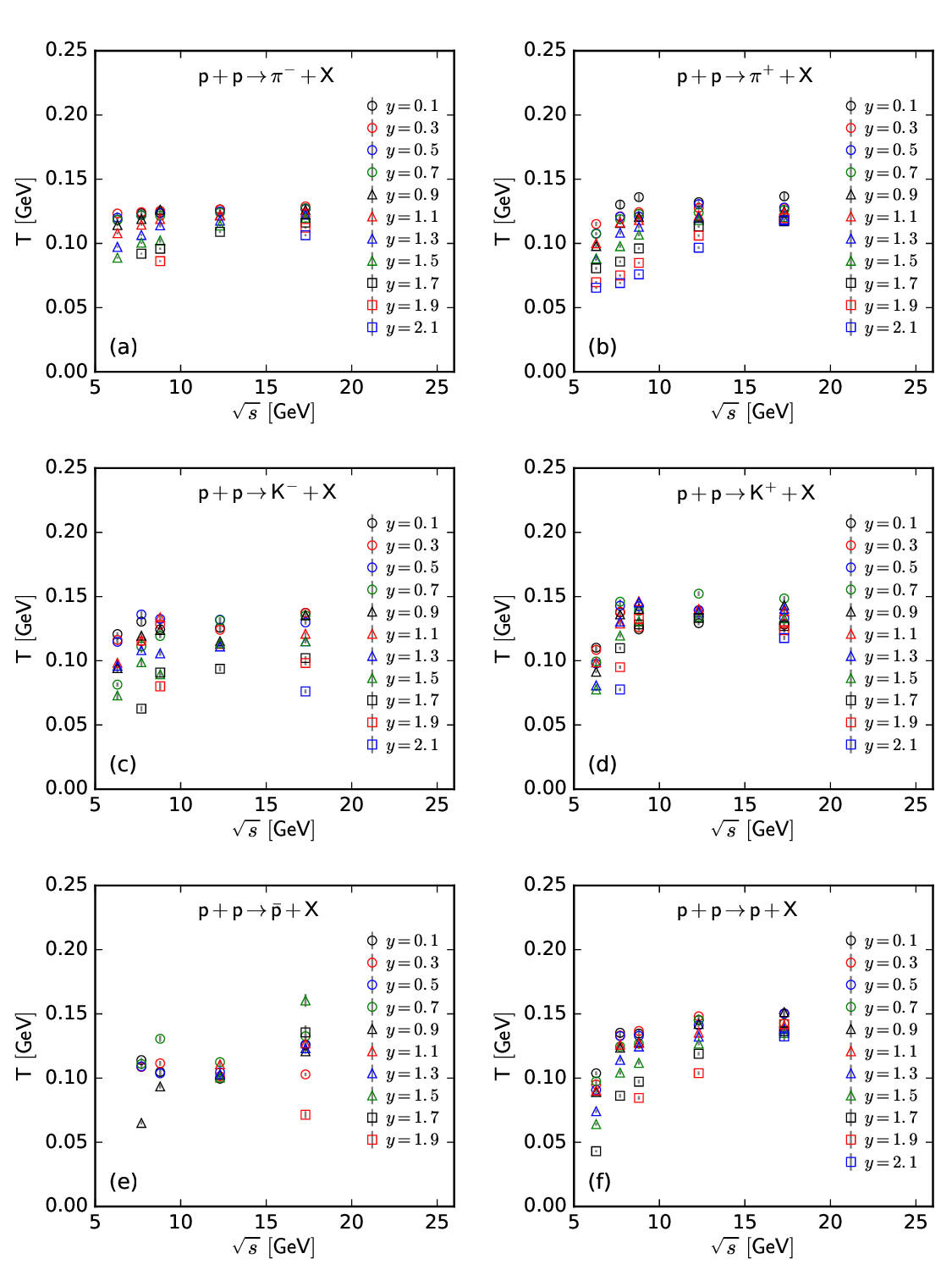}
\vskip0.2cm \justifying\noindent {Fig. 11. Dependence of $T$ on
$\sqrt{s}$ at different $y$ from the spectra of (a) $\pi^{-}$, (b)
$\pi^{+}$, (c) $K^{-}$, (d) $K^{+}$, (e) $\bar{p}$, and (f) $p$.}
\end{figure*}

\begin{figure*}[htbp]
\vskip-1.5cm \hspace{0.cm}
\includegraphics[width=17.25cm]{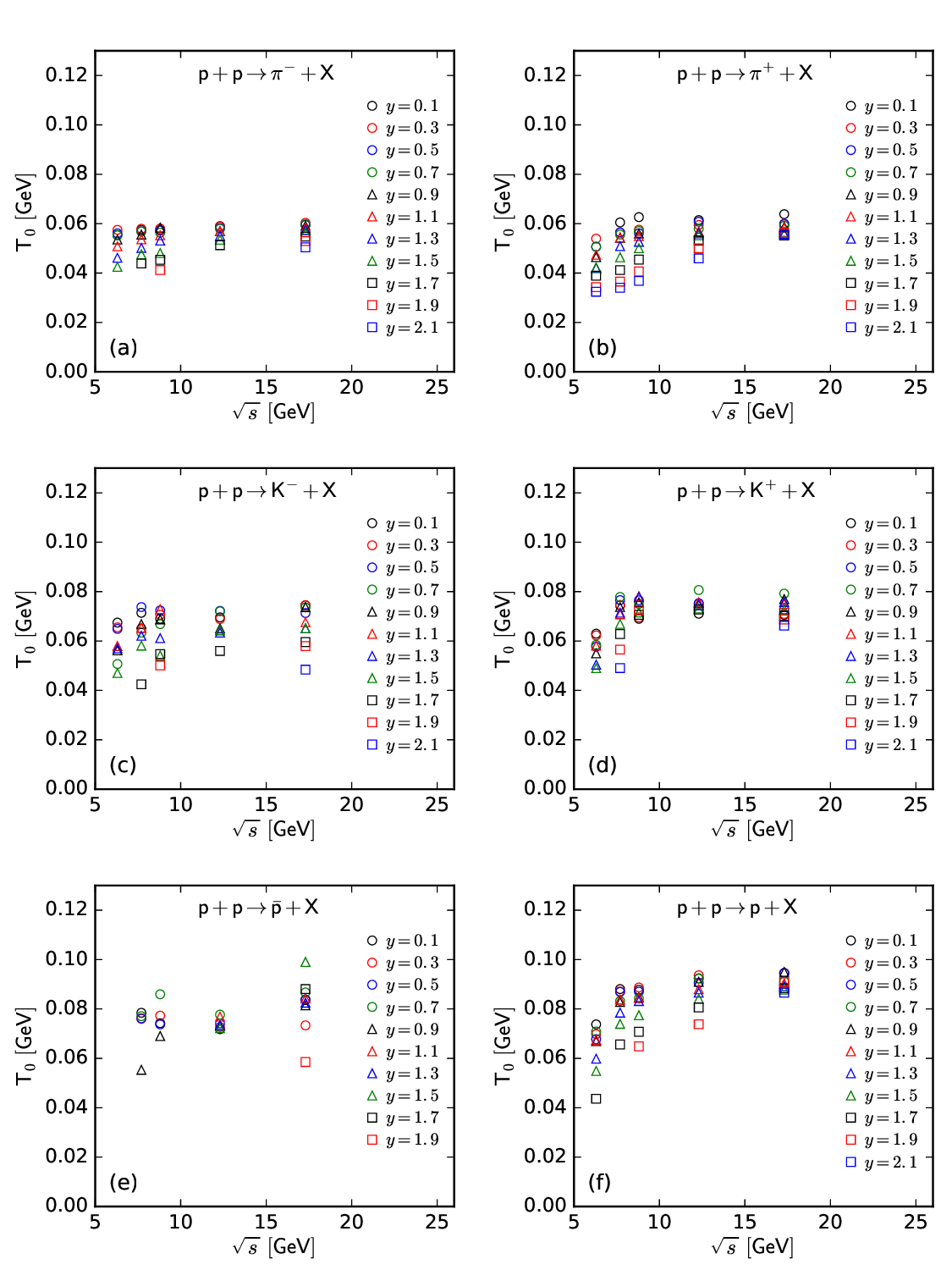}
\vskip0.2cm \justifying\noindent {Fig. 12. Same as Figure 11, but
for the dependence of $T_0$ on $\sqrt{s}$ at different $y$.}
\end{figure*}

\begin{figure*}[htbp]
\vskip-1.5cm \hspace{0.cm}
\includegraphics[width=17.25cm]{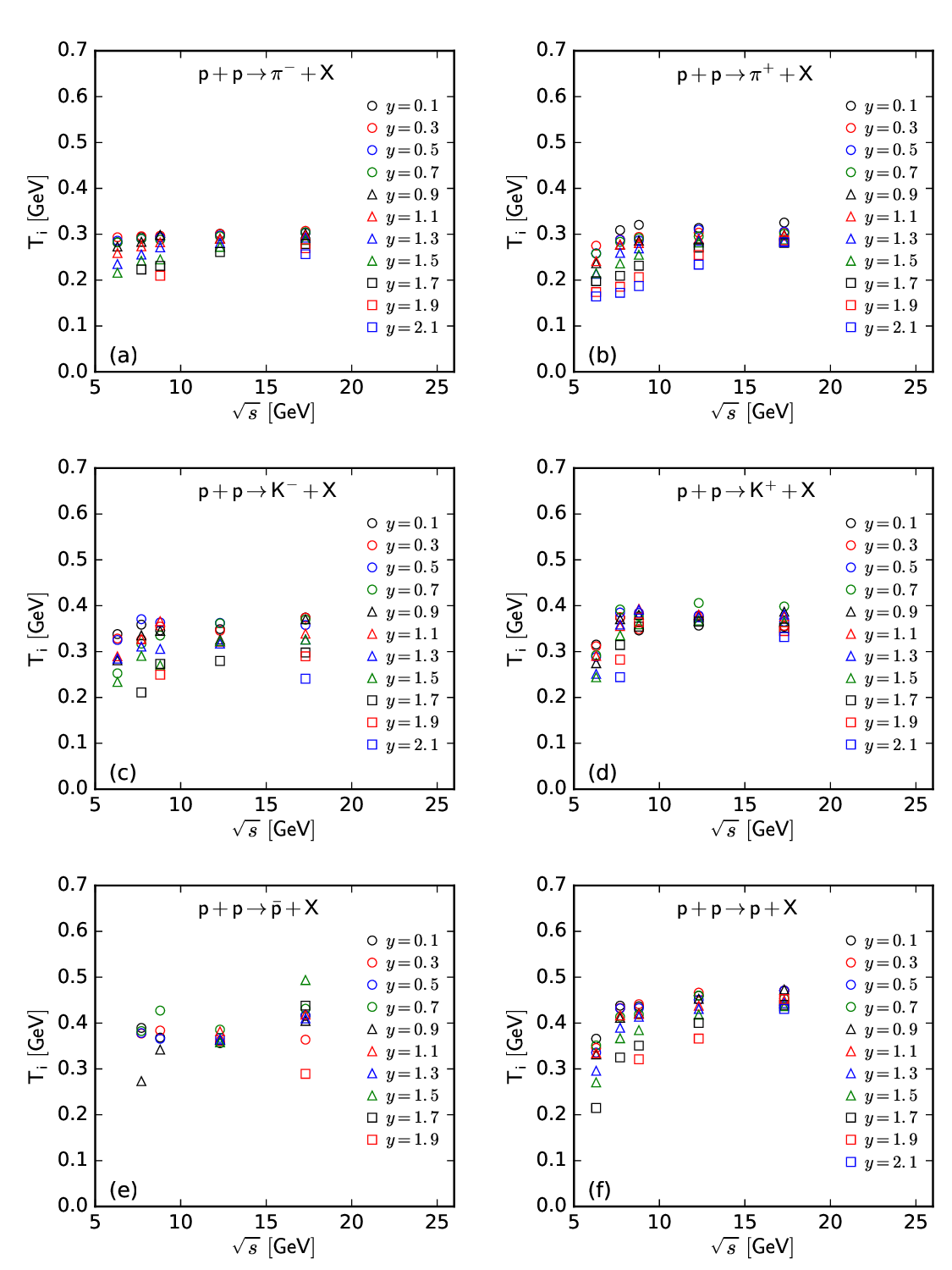}
\vskip0.2cm \justifying\noindent {Fig. 13. Same as Figure 11, but
for the dependence of $T_i$ on $\sqrt{s}$ at different $y$.}
\end{figure*}

\begin{figure*}[htbp]
\vskip-1.5cm \hspace{0.cm}
\includegraphics[width=17.25cm]{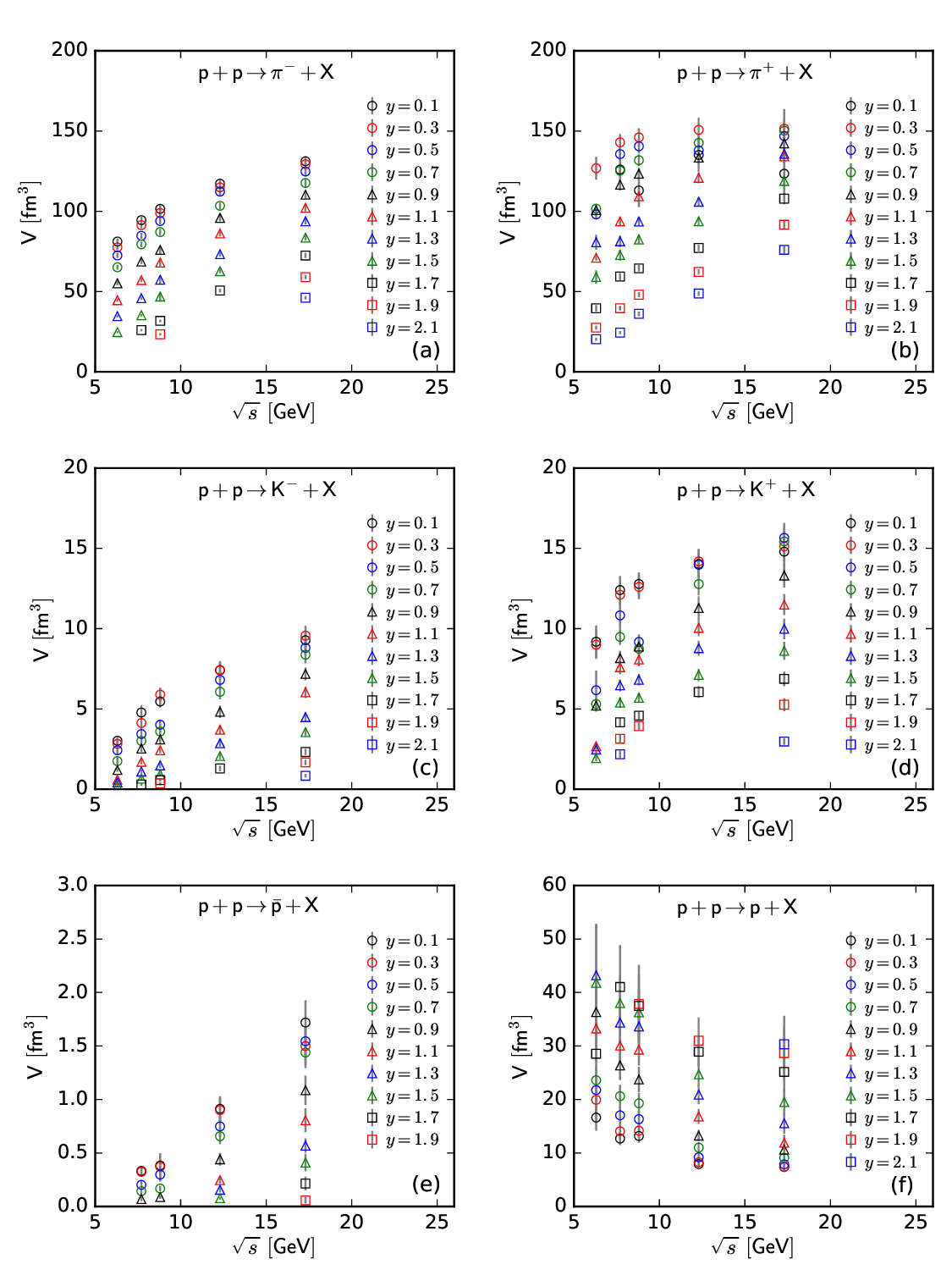}
\vskip0.2cm \justifying\noindent {Fig. 14. Same as Figure 11, but
for the dependence of $V$ on $\sqrt{s}$ at different $y$.}
\end{figure*}

In the above discussions, although chemical potential $\mu$ runs
through the entire process, it is an insensitive quantity in the
fit, and not a free parameter due to the fact that it depends on
$T_{ch}$ and $k_j$. Our previous work~\cite{48c,48d} shows that,
from 6.3 to 17.3 GeV, $\mu_{\pi^+}$, $\mu_{K^+}$, and $\mu_p$ are
around 0.041--0.017, 0.110--0.042, and 0.510--0.180 GeV,
respectively, which are directly used in this work. These results
are excluded the contributions from resonance decays~\cite{56-a}.
Although the resonance decays contribute considerably to the
yields of negative and positive hadrons, they contribute to the
yield ratios and then the chemical potentials being
small~\cite{48c,48d}.

Before summary and conclusions, it should be pointed out that the
data sets analyzed by us are in a narrow and low-$p_T$ range,
which obey the standard distribution. We believe that even if the
narrow spectra are in a high-$p_T$ range, the standard
distribution can be used and a high temperature can be obtained.
The success of this work reflects that the classical concept and
distribution can still play a great role in the field of high
energy collisions, though the application is in a local region. In
our opinion, when researchers search for novel theoretical models,
they first need to take into account classical theories.

Although the topic has been extensively studied in many papers for
the SPS, RHIC and LHC heavy ion collisions and outlined the
validity of a nonextensive statistical
distribution~\cite{56,57,58,59,60,61,62}, those investigations
used the spectra in a wide $p_T$ range. It is unanimous that for
the wide $p_T$ spectra, a two-, three-, or multi-component
standard distribution is needed in the fit. Then, a temperature
fluctuation can be observed from the multi-component standard
distribution. At this point, the Tsallis distribution is needed.
This is the relationship between the standard distribution and the
Tsallis distribution in the fit process.

In addition, in comparison with Hanbury-Brown-Twiss (HBT)
results~\cite{63}, large values of volume are obtained in the
present work. The reason is that different volumes are studied.
Generally, the former describes the system size in the initial
state of collisions, and the latter is a reflection of the size of
an expanded fireball in the final state (at the kinetic
freeze-out) of collisions. Obviously, the latter is much larger
than the former. The values of three temperatures obtained in the
present work seem reasonable.

\section{Summary and conclusions}

The main observations and conclusions are summarized here.

(a) The transverse momentum spectra of identified charged hadrons
($\pi^{-}$, $\pi^{+}$, $K^{-}$, $K^{+}$, $\bar{p}$, and $p$) with
different rapidities produced in proton-proton collisions at
center-of-mass energies $\sqrt{s}=6.3$, 7.7, 8.8, 12.3, and 17.3
GeV have been studied using the standard distribution. The fitted
results are in agreement with the experimental data measured by
the NA61/SHINE Collaboration at the SPS. The effective temperature
$T$, kinetic freeze-out temperature $T_{0}$, initial temperature
$T_{i}$, and kinetic freeze-out volume $V$ are extracted. The
present work shows that the standard distribution coming from the
relativistic ideal gas model works well in some cases.

(b) In most cases, $T$, $T_0$, and $T_i$ decrease with the
increase in rapidity $y$, and increase with the increase in
$\sqrt{s}$. There is a tendency of saturation for the three
temperatures at $\sqrt{s}=7.7$ GeV and above. From a quick
increase to a slow saturation in the three temperatures, the
transition energy 7.7 GeV is the boundary for proton-dominated and
meson-dominated final states. For the spectra of produced hadrons
($\pi^{-}$, $\pi^{+}$, $K^{-}$, $K^{+}$, and $\bar{p}$), the
extracted $V$ also decreases with the increase in $y$, and
increases with the increase in $\sqrt{s}$. For the spectra of $p$,
the extracted $V$ increases with the increase in $y$ and decreases
with the increase in $\sqrt{s}$. This is opposite to other
hadrons, because $p$ contains the pre-existing leading protons
which affect the result.

(c) The three temperatures do not show an obvious isospin
dependence. However, $V$ shows a significant isospin dependence.
The reason for the isospin dependence of $V$ is the
electromagnetic interactions between positive (negative) hadrons
and pre-existing protons. The exclusion (attraction) between
positive (negative) hadrons and pre-existing protons causes larger
(smaller) $V$ of emission source of positive (negative) hadrons.
Compared with the three temperature types, $V$ shows a larger
mass-dependence. The mass-dependence of $V$ is also a reflection
of a mass-dependent differential kinetic freeze-out scenario or
multiple kinetic freeze-out scenario.
\\
\\
{\bf Author Contributions:} Conceptualization, F.-H.L. and K.K.O.;
Methodology, F.-H.L. and K.K.O.; Software, P.-P.Y.; Validation,
F.-H.L. and K.K.O.; Formal analysis, P.-P.Y.; Investigation,
P.-P.Y.; Resources, P.-P.Y.; Data curation, P.-P.Y.; Writing --
original draft, P.-P.Y.; Writing -- review and editing, F.-H.L.
and K.K.O.; Visualization, P.-P.Y.; Supervision, F.-H.L. and
K.K.O.; Project administration, P.-P.Y. and F.-H.L.; Funding
acquisition, P.-P.Y., F.-H.L. and K.K.O. All authors have read and
agreed to the published version of the manuscript.
\\
\\
{\bf Funding:} The work of P.-P.Y. was supported by the Shanxi
Provincial Natural Science Foundation under Grant No.
202203021222308 and the Doctoral Scientific Research Foundations
of Shanxi Province and Xinzhou Normal University. The work of
F.-H.L. was supported by the National Natural Science Foundation
of China under Grant No. 12147215, the Shanxi Provincial Natural
Science Foundation under Grant No. 202103021224036, and the Fund
for Shanxi ``1331 Project" Key Subjects Construction. The work of
K.K.O. was supported by the Agency of Innovative Development under
the Ministry of Higher Education, Science and Innovations of the
Republic of Uzbekistan within the fundamental project No.
F3-20200929146 on analysis of open data on heavy-ion collisions at
RHIC and LHC.
\\
\\
{\bf Institutional Review Board Statement:} Not applicable.
\\
\\
{\bf Informed Consent Statement:} Not applicable.
\\
\\
{\bf Data Availability Statement:} The data used to support the
findings of this study are included within the article and are
cited at relevant places within the text as references.
\\
\\
{\bf Conflict of Interest:} The authors declare no conflict of
interest. The funding agencies have no role in the design of the
study; in the collection, analysis, or interpretation of the data;
in the writing of the manuscript, or in the decision to publish
the results.
\\

\vskip1.0cm

\noindent {\bf Appendix A: The method to obtain the parameter and
its uncertainty}
\\

Let
\begin{align}
y_{i}=f(x_{i}), \quad i=1,2,...,n
\end{align}
be the model value of the $i$-th fitting point. The physical
quantities or parameters, \\ $\lambda \ (\lambda_{1}, \lambda_{2},
..., \lambda_{j}$), can be obtained by fitting the experimental
data, where $j$ is the number of parameters which includes the
normalization constant. One has
\begin{align}
\chi^2=\sum^{n}_{i=1}\frac{[f(x_{i})-Y_{i}]^2}{(\delta Y_{i})^2},
\end{align}
where $n$ is the number of fitting points, $Y_{i}$ represent the
experimental value and $\delta Y_{i}$ represent the uncertainty of
experimental value, usually including statistical and systematic
uncertainties.

Due to the small particle number of $p_T$ samples being studied in
this paper, the parameter uncertainty is assumed to follow the
Student's distribution (shorted to t-distribution)~\cite{54},
\begin{align}
f(\lambda\mid\nu)=\frac{(\frac{\nu}{\nu+\lambda^2})^{\frac{\nu+1}{2}}\Gamma(\frac{\nu+1}{2})}{\sqrt{\mu
\pi }\Gamma(\frac{\nu}{2})},
\end{align}
where $\nu$ represent the ndof, $\Gamma(x)$ represent the Gamma
function. With the increase in $\nu$, the t-distribution gradually
approaches to the normal or Gaussian distribution, $N(0,1)$. When
$\nu$ approaches 1, t-distribution approaches the Cauchy
distribution.

In the present fitting, a $0.5\%$ confidence interval is used to
describe the parameter uncertainty. This means that there is a
$0.5\%$ probability that the parameter will fall between
($\lambda_{j}-t\sigma, \lambda_{j}+t\sigma$), where $\sigma$ is
the standard deviation of each parameter and $t$ satisfies the
following equation,
\begin{align}
P=\int^{t\sigma}_{-t\sigma}f(\lambda)d\lambda=0.005.
\end{align}

The standard deviation of each parameter can be calculated by
\begin{align}
\sigma=\sqrt{(J_{\lambda}^{T}J_{\lambda})^{-1}\frac{s^2}{\nu}},
\end{align}
where
\begin{align}
(J_{\lambda})_{ij}=\frac{\partial [f(x_{i})-Y_{i}]}{\partial
\lambda_{j}}
\end{align}
is the Jacobian matrix and determinant of the model,
$J_{\lambda}^{T}$ is the transpose of Jacobian matrix, the
superscript $-1$ represents matrix inversion,
\begin{align}
s^2= \sum_{i=1}^{n}[f(x_{i})-Y_{i}]^2
\end{align}
is the variance. Then, the corresponding best parameter is given
by
\begin{align}
\lambda_{j}\in[\lambda_{j}-t\sigma_{j,j},\lambda_{j}+t\sigma_{j,j}]
\end{align}
with its uncertainty of $t\sigma_{j,j}$.
\\
\\
\noindent {\bf Appendix B: Parameter tables obtained in the
fitting process}

\begin{table*}[htbp] \vspace{0.cm} \justifying\noindent {\small Table 1.
Values of $T$, $V$, $\chi^2$, and ndof corresponding to the curves
in Figures 1 and 2 for $\pi^{-}$ and $\pi^+$ produced in $pp$
collisions respectively, where the values of $\sqrt{s}$ is given
together. The values of $\chi^2$ are reserved as an integer or the
first non-zero decimal (if less than 1). The symbols ``$-$"
indicate that the data are not available.\\} \vspace{-0.5cm}
{\footnotesize \hspace{0.cm}\begin{center}
\newcommand{\tabincell}[2]{\begin{tabular}{@{}#1@{}}#2\end{tabular}}
\begin{tabular} {cc|ccc|ccc}\\ \hline
$\sqrt{s}$ (GeV) & $y$ & \tabincell{c}{\\$T$ (GeV)} & \tabincell{c}{$\pi^-$\\$V$ (fm$^3$)} &\tabincell{c}{\\$\chi^2$/ndof} & \tabincell{c}{\\$T$ (GeV)} & \tabincell{c}{$\pi^+$\\$V$ (fm$^3$)} & \tabincell{c}{\\$\chi^2$/ndof} \\
\hline
      & $0.1$ & $0.118\pm0.001$ & $(8.119\pm0.177)\times10^{1}$ & $13/16$ & $-$             & $-$                           & $-$\\
      & $0.3$ & $0.123\pm0.001$ & $(7.755\pm0.167)\times10^{1}$ & $17/16$ & $0.115\pm0.002$ & $(1.268\pm0.072)\times10^{2}$ & $0.09/6$\\
      & $0.5$ & $0.119\pm0.001$ & $(7.247\pm0.157)\times10^{1}$ & $15/16$ & $0.108\pm0.001$ & $(9.805\pm0.269)\times10^{1}$ & $2/10$\\
      & $0.7$ & $0.118\pm0.001$ & $(6.514\pm0.141)\times10^{1}$ & $11/16$ & $0.107\pm0.001$ & $(1.017\pm0.028)\times10^{2}$ & $1/10$\\
      & $0.9$ & $0.114\pm0.001$ & $(5.516\pm0.121)\times10^{1}$ & $11/16$ & $0.098\pm0.001$ & $(1.007\pm0.029)\times10^{2}$ & $3/10$\\
6.3   & $1.1$ & $0.108\pm0.001$ & $(4.464\pm0.098)\times10^{1}$ & $16/16$ & $0.100\pm0.001$ & $(7.106\pm0.179)\times10^{1}$ & $2/9$\\
      & $1.3$ & $0.097\pm0.001$ & $(3.466\pm0.081)\times10^{1}$ & $14/15$ & $0.087\pm0.001$ & $(8.072\pm0.468)\times10^{1}$ & $1/6$\\
      & $1.5$ & $0.088\pm0.001$ & $(2.472\pm0.061)\times10^{1}$ & $9/14$  & $0.088\pm0.001$ & $(5.903\pm0.423)\times10^{1}$ & $1/4$\\
      & $1.7$ & $-$             & $-$                           & $-$     & $0.081\pm0.001$ & $(3.957\pm0.196)\times10^{1}$ & $4/5$\\
      & $1.9$ & $-$             & $-$                           & $-$     & $0.070\pm0.001$ & $(2.736\pm0.112)\times10^{1}$ & $4/6$\\
      & $2.1$ & $-$             & $-$                           & $-$     & $0.066\pm0.001$ & $(2.036\pm0.099)\times10^{1}$ & $0.1/5$\\
\hline
      & $0.1$ & $0.123\pm0.001$ & $(9.450\pm0.204)\times10^{1}$ & $43/16$ & $0.130\pm0.003$ & $(1.261\pm0.137)\times10^{2}$ & $0.5/3$\\
      & $0.3$ & $0.124\pm0.001$ & $(9.141\pm0.197)\times10^{1}$ & $33/16$ & $0.121\pm0.001$ & $(1.423\pm0.052)\times10^{2}$ & $0.5/6$\\
      & $0.5$ & $0.122\pm0.001$ & $(8.475\pm0.183)\times10^{1}$ & $32/16$ & $0.121\pm0.001$ & $(1.356\pm0.037)\times10^{2}$ & $2/9$\\
      & $0.7$ & $0.122\pm0.001$ & $(7.944\pm0.172)\times10^{1}$ & $19/16$ & $0.119\pm0.001$ & $(1.253\pm0.035)\times10^{2}$ & $1/9$\\
      & $0.9$ & $0.119\pm0.001$ & $(6.865\pm0.149)\times10^{1}$ & $20/16$ & $0.116\pm0.001$ & $(1.165\pm0.032)\times10^{2}$ & $2/9$\\
7.7   & $1.1$ & $0.115\pm0.001$ & $(5.715\pm0.124)\times10^{1}$ & $17/16$ & $0.116\pm0.001$ & $(9.373\pm0.251)\times10^{1}$ & $0.5/8$\\
      & $1.3$ & $0.107\pm0.001$ & $(4.585\pm0.104)\times10^{1}$ & $14/15$ & $0.108\pm0.001$ & $(8.131\pm0.354)\times10^{1}$ & $1/7$\\
      & $1.5$ & $0.101\pm0.001$ & $(3.526\pm0.081)\times10^{1}$ & $12/15$ & $0.098\pm0.001$ & $(7.273\pm0.343)\times10^{1}$ & $1/7$\\
      & $1.7$ & $0.092\pm0.001$ & $(2.600\pm0.066)\times10^{1}$ & $12/13$ & $0.086\pm0.001$ & $(5.933\pm0.223)\times10^{1}$ & $4/8$\\
      & $1.9$ & $-$             & $-$                           & $-$     & $0.075\pm0.001$ & $(3.963\pm0.141)\times10^{1}$ & $12/8$\\
      & $2.1$ & $-$             & $-$                           & $-$     & $0.069\pm0.001$ & $(2.440\pm0.115)\times10^{1}$ & $35/9$\\
\hline
      & $0.1$ & $0.124\pm0.001$ & $(1.015\pm0.025)\times10^{2}$ & $53/16$ & $0.136\pm0.003$ & $(1.130\pm0.104)\times10^{2}$ & $0.2/4$\\
      & $0.3$ & $0.125\pm0.001$ & $(9.894\pm0.219)\times10^{1}$ & $70/16$ & $0.124\pm0.001$ & $(1.460\pm0.057)\times10^{2}$ & $1/7$\\
      & $0.5$ & $0.124\pm0.001$ & $(9.394\pm0.208)\times10^{1}$ & $69/16$ & $0.121\pm0.001$ & $(1.405\pm0.044)\times10^{2}$ & $1/8$\\
      & $0.7$ & $0.122\pm0.001$ & $(8.704\pm0.193)\times10^{1}$ & $43/16$ & $0.123\pm0.001$ & $(1.318\pm0.036)\times10^{2}$ & $1/9$\\
      & $0.9$ & $0.126\pm0.001$ & $(7.596\pm0.245)\times10^{1}$ & $2/7$   & $0.121\pm0.001$ & $(1.235\pm0.035)\times10^{2}$ & $1/9$\\
8.8   & $1.1$ & $0.119\pm0.001$ & $(6.808\pm0.245)\times10^{1}$ & $0.2/6$ & $0.118\pm0.001$ & $(1.093\pm0.031)\times10^{2}$ & $0.4/9$\\
      & $1.3$ & $0.114\pm0.001$ & $(5.736\pm0.229)\times10^{1}$ & $1/5$   & $0.113\pm0.001$ & $(9.364\pm0.273)\times10^{1}$ & $1/9$\\
      & $1.5$ & $0.102\pm0.001$ & $(4.687\pm0.238)\times10^{1}$ & $0.2/4$ & $0.107\pm0.001$ & $(8.245\pm0.270)\times10^{1}$ & $0.2/8$\\
      & $1.7$ & $0.096\pm0.001$ & $(3.175\pm0.076)\times10^{1}$ & $11/15$ & $0.096\pm0.001$ & $(6.446\pm0.198)\times10^{1}$ & $4/9$\\
      & $1.9$ & $0.086\pm0.001$ & $(2.337\pm0.062)\times10^{1}$ & $8/13$  & $0.084\pm0.001$ & $(4.803\pm0.152)\times10^{1}$ & $4/9$\\
      & $2.1$ & $-$             & $-$                           & $-$     & $0.076\pm0.001$ & $(3.612\pm0.137)\times10^{1}$ & $3/7$\\
\hline
      & $0.1$ & $0.126\pm0.001$ & $(1.172\pm0.025)\times10^{2}$ & $88/16$ & $0.132\pm0.002$ & $(1.351\pm0.106)\times10^{2}$ & $0.3/4$\\
      & $0.3$ & $0.126\pm0.001$ & $(1.149\pm0.025)\times10^{2}$ & $91/16$ & $0.128\pm0.002$ & $(1.507\pm0.076)\times10^{2}$ & $0.4/6$\\
      & $0.5$ & $0.125\pm0.001$ & $(1.123\pm0.024)\times10^{2}$ & $69/16$ & $0.131\pm0.001$ & $(1.379\pm0.046)\times10^{2}$ & $2/8$\\
      & $0.7$ & $0.124\pm0.001$ & $(1.034\pm0.022)\times10^{2}$ & $60/16$ & $0.124\pm0.001$ & $(1.426\pm0.046)\times10^{2}$ & $0.4/7$\\
      & $0.9$ & $0.122\pm0.001$ & $(9.590\pm0.206)\times10^{1}$ & $51/16$ & $0.121\pm0.001$ & $(1.334\pm0.039)\times10^{2}$ & $1/8$\\
12.3  & $1.1$ & $0.122\pm0.001$ & $(8.621\pm0.185)\times10^{1}$ & $43/16$ & $0.122\pm0.001$ & $(1.208\pm0.033)\times10^{2}$ & $1/9$\\
      & $1.3$ & $0.118\pm0.001$ & $(7.327\pm0.158)\times10^{1}$ & $27/16$ & $0.120\pm0.001$ & $(1.060\pm0.026)\times10^{2}$ & $1/11$\\
      & $1.5$ & $0.114\pm0.001$ & $(6.270\pm0.136)\times10^{1}$ & $19/16$ & $0.118\pm0.001$ & $(9.376\pm0.233)\times10^{1}$ & $1/11$\\
      & $1.7$ & $0.109\pm0.001$ & $(5.063\pm0.112)\times10^{1}$ & $20/16$ & $0.113\pm0.001$ & $(7.718\pm0.020)\times10^{1}$ & $2/11$\\
      & $1.9$ & $-$             & $-$                           & $-$     & $0.106\pm0.001$ & $(6.229\pm0.016)\times10^{1}$ & $3/11$\\
      & $2.1$ & $-$             & $-$                           & $-$     & $0.097\pm0.001$ & $(4.876\pm0.014)\times10^{1}$ & $4/10$\\
\hline
      & $0.1$ & $0.127\pm0.001$ & $(1.312\pm0.028)\times10^{2}$ & $89/16$ & $0.137\pm0.003$ & $(1.234\pm0.145)\times10^{2}$ & $0.05/3$\\
      & $0.3$ & $0.129\pm0.001$ & $(1.293\pm0.027)\times10^{2}$ & $99/16$ & $0.127\pm0.002$ & $(1.514\pm0.127)\times10^{2}$ & $0.1/4$\\
      & $0.5$ & $0.127\pm0.001$ & $(1.248\pm0.026)\times10^{2}$ & $93/16$ & $0.128\pm0.002$ & $(1.469\pm0.087)\times10^{2}$ & $0.4/5$\\
      & $0.7$ & $0.127\pm0.001$ & $(1.176\pm0.025)\times10^{2}$ & $92/16$ & $0.126\pm0.002$ & $(1.502\pm0.064)\times10^{2}$ & $1/6$\\
      & $0.9$ & $0.126\pm0.001$ & $(1.104\pm0.023)\times10^{2}$ & $87/16$ & $0.125\pm0.002$ & $(1.442\pm0.061)\times10^{2}$ & $2/6$\\
17.3  & $1.1$ & $0.124\pm0.001$ & $(1.022\pm0.021)\times10^{2}$ & $66/16$ & $0.125\pm0.001$ & $(1.342\pm0.045)\times10^{2}$ & $1/7$\\
      & $1.3$ & $0.122\pm0.001$ & $(9.378\pm0.971)\times10^{1}$ & $37/16$ & $0.120\pm0.001$ & $(1.358\pm0.045)\times10^{2}$ & $1/7$\\
      & $1.5$ & $0.120\pm0.001$ & $(8.359\pm0.177)\times10^{1}$ & $27/16$ & $0.120\pm0.001$ & $(1.189\pm0.039)\times10^{2}$ & $0.2/7$\\
      & $1.7$ & $0.116\pm0.001$ & $(7.247\pm0.154)\times10^{1}$ & $20/16$ & $0.117\pm0.001$ & $(1.079\pm0.035)\times10^{2}$ & $0.4/7$\\
      & $1.9$ & $0.112\pm0.001$ & $(5.898\pm0.126)\times10^{1}$ & $16/16$ & $0.120\pm0.001$ & $(9.162\pm0.030)\times10^{1}$ & $0.1/7$\\
      & $2.1$ & $0.106\pm0.001$ & $(4.619\pm0.100)\times10^{1}$ & $20/16$ & $0.118\pm0.001$ & $(7.601\pm0.024)\times10^{1}$ & $0.2/7$\\
\hline
\end{tabular}%
\end{center}}
\end{table*}

\begin{table*}[htbp] \vspace{0.cm} \justifying\noindent {\small Table 2. Same
as Table 1, but corresponding to the curves in Figures 3 and 4 for
$K^-$ and $K^+$ respectively.\\} \vspace{-0.5cm} {\small
\hspace{0.cm} \begin{center}
\newcommand{\tabincell}[2]{\begin{tabular}{@{}#1@{}}#2\end{tabular}}
\begin{tabular} {cc|ccc|ccc}\\ \hline
$\sqrt{s}$ (GeV) & $y$ & \tabincell{c}{\\$T$ (GeV)} & \tabincell{c}{$K^-$\\$V$ (fm$^3$)} &\tabincell{c}{\\$\chi^2$/ndof} & \tabincell{c}{\\$T$ (GeV)} & \tabincell{c}{$K^+$\\$V$ (fm$^3$)} & \tabincell{c}{\\$\chi^2$/ndof} \\
\hline
      & $0.1$ & $0.121\pm0.001$ & $(3.013\pm0.186)\times10^{0}$  & $2/8$   & $0.110\pm0.002$ & $(9.184\pm0.990)\times10^0$ & $1/5$\\
      & $0.3$ & $0.116\pm0.001$ & $(2.824\pm0.195)\times10^{0}$  & $1/7$   & $0.108\pm0.002$ & $(9.001\pm0.863)\times10^0$ & $3/6$\\
      & $0.5$ & $0.115\pm0.002$ & $(2.422\pm0.284)\times10^{0}$  & $2/4$   & $0.098\pm0.003$ & $(6.165\pm1.215)\times10^0$ & $6/4$\\
6.3   & $0.7$ & $0.081\pm0.001$ & $(1.748\pm0.247)\times10^{0}$  & $3/4$   & $0.100\pm0.001$ & $(5.302\pm0.461)\times10^0$ & $18/10$\\
      & $0.9$ & $0.094\pm0.001$ & $(1.190\pm0.108)\times10^{0}$  & $4/6$   & $0.091\pm0.001$ & $(5.201\pm0.359)\times10^0$ & $5/10$\\
      & $1.1$ & $0.098\pm0.002$ & $(5.859\pm0.722)\times10^{-1}$ & $4/4$   & $0.099\pm0.001$ & $(2.675\pm0.166)\times10^0$ & $10/9$\\
      & $1.3$ & $0.096\pm0.005$ & $(4.229\pm1.372)\times10^{-1}$ & $0.09/2$& $0.081\pm0.001$ & $(2.467\pm0.243)\times10^0$ & $3/6$\\
      & $1.5$ & $0.073\pm0.003$ & $(2.041\pm0.654)\times10^{-1}$ & $0.7/2$ & $0.078\pm0.001$ & $(1.935\pm0.200)\times10^0$ & $5/6$\\
\hline
      & $0.1$ & $0.130\pm0.002$ & $(4.775\pm0.457)\times10^{0}$  & $0.5/5$ & $0.137\pm0.002$ & $(1.240\pm0.087)\times10^{1}$ & $1/7$\\
      & $0.3$ & $0.115\pm0.002$ & $(4.123\pm0.335)\times10^{0}$  & $2/6$   & $0.138\pm0.002$ & $(1.211\pm0.065)\times10^{1}$ & $1/9$\\
      & $0.5$ & $0.136\pm0.003$ & $(3.434\pm0.386)\times10^{0}$  & $1/4$   & $0.142\pm0.001$ & $(1.083\pm0.084)\times10^{1}$ & $1/6$\\
      & $0.7$ & $0.111\pm0.002$ & $(3.004\pm0.244)\times10^{0}$  & $1/6$   & $0.146\pm0.001$ & $(9.481\pm0.497)\times10^0$   & $1/9$\\
      & $0.9$ & $0.120\pm0.002$ & $(2.526\pm0.197)\times10^{0}$  & $1/6$   & $0.136\pm0.001$ & $(8.165\pm0.439)\times10^0$   & $1/9$\\
7.7   & $1.1$ & $0.116\pm0.002$ & $(1.703\pm0.162)\times10^{0}$  & $1/5$   & $0.129\pm0.001$ & $(7.598\pm0.423)\times10^0$   & $1/9$\\
      & $1.3$ & $0.108\pm0.002$ & $(1.097\pm0.108)\times10^{0}$  & $1/5$   & $0.130\pm0.001$ & $(6.466\pm0.358)\times10^0$   & $4/9$\\
      & $1.5$ & $0.099\pm0.003$ & $(6.563\pm1.121)\times10^{-1}$ & $0.3/3$ & $0.120\pm0.001$ & $(5.394\pm0.317)\times10^0$   & $2/9$\\
      & $1.7$ & $0.063\pm0.002$ & $(2.875\pm0.807)\times10^{-1}$ & $0.4/2$ & $0.110\pm0.001$ & $(4.168\pm0.277)\times10^0$   & $1/8$\\
      & $1.9$ & $-$             & $-$                            & $-$     & $0.095\pm0.001$ & $(3.140\pm0.256)\times10^0$   & $2/7$\\
      & $2.1$ & $-$             & $-$                            & $-$     & $0.078\pm0.001$ & $(2.172\pm0.229)\times10^0$   & $7/6$\\
\hline
      & $0.1$ & $0.124\pm0.002$ & $(5.460\pm0.357)\times10^{0}$  & $3/7$   & $0.124\pm0.001$ & $(1.278\pm0.072)\times10^{1}$ & $3/9$\\
      & $0.3$ & $0.128\pm0.002$ & $(5.890\pm0.431)\times10^{0}$  & $0.4/6$ & $0.126\pm0.001$ & $(1.259\pm0.076)\times10^{1}$ & $1/8$\\
      & $0.5$ & $0.133\pm0.002$ & $(4.015\pm0.250)\times10^{0}$  & $3/7$   & $0.142\pm0.001$ & $(9.172\pm0.468)\times10^0$   & $2/9$\\
      & $0.7$ & $0.119\pm0.002$ & $(3.591\pm0.245)\times10^{0}$  & $1/7$   & $0.139\pm0.001$ & $(8.748\pm0.463)\times10^0$   & $1/9$\\
8.8   & $0.9$ & $0.124\pm0.002$ & $(3.099\pm0.239)\times10^{0}$  & $0.4/6$ & $0.140\pm0.001$ & $(8.885\pm0.470)\times10^0$   & $0.5/9$\\
      & $1.1$ & $0.134\pm0.002$ & $(2.426\pm0.185)\times10^{0}$  & $1/6$   & $0.146\pm0.002$ & $(8.079\pm0.418)\times10^0$   & $1/9$\\
      & $1.3$ & $0.106\pm0.002$ & $(1.478\pm0.144)\times10^{0}$  & $1/5$   & $0.146\pm0.001$ & $(6.816\pm0.352)\times10^0$   & $1/9$\\
      & $1.5$ & $0.089\pm0.002$ & $(9.181\pm1.199)\times10^{-1}$ & $0.2/4$ & $0.131\pm0.001$ & $(5.690\pm0.313)\times10^0$   & $1/9$\\
      & $1.7$ & $0.091\pm0.002$ & $(5.613\pm0.733)\times10^{-1}$ & $2/4$   & $0.128\pm0.001$ & $(4.577\pm0.252)\times10^0$   & $2/9$\\
      & $1.9$ & $0.080\pm0.003$ & $(3.513\pm0.860)\times10^{-1}$ & $1/2$   & $0.132\pm0.001$ & $(3.929\pm0.228)\times10^0$   & $5/8$\\
\hline
      & $0.1$ & $0.125\pm0.002$ & $(7.420\pm0.487)\times10^0$ & $1/7$    & $0.129\pm0.002$ & $(1.404\pm0.093)\times10^{1}$ & $1/7$\\
      & $0.3$ & $0.124\pm0.002$ & $(7.399\pm0.563)\times10^0$ & $0.3/6$  & $0.139\pm0.002$ & $(1.419\pm0.075)\times10^{1}$ & $2/9$\\
      & $0.5$ & $0.132\pm0.002$ & $(6.808\pm0.504)\times10^0$ & $0.4/6$  & $0.139\pm0.001$ & $(1.398\pm0.073)\times10^{1}$ & $1/9$\\
      & $0.7$ & $0.132\pm0.002$ & $(6.064\pm0.452)\times10^0$ & $0.4/6$  & $0.152\pm0.002$ & $(1.278\pm0.070)\times10^{1}$ & $0.1/8$\\
12.3  & $0.9$ & $0.115\pm0.001$ & $(4.823\pm0.380)\times10^0$ & $0.5/6$  & $0.139\pm0.002$ & $(1.128\pm0.072)\times10^{1}$ & $1/7$\\
      & $1.1$ & $0.113\pm0.001$ & $(3.704\pm0.262)\times10^0$ & $1/7$    & $0.140\pm0.002$ & $(1.005\pm0.523)\times10^{1}$ & $0.4/9$\\
      & $1.3$ & $0.111\pm0.001$ & $(2.861\pm0.207)\times10^0$ & $0.4/7$  & $0.137\pm0.001$ & $(8.783\pm0.463)\times10^0$   & $1/9$\\
      & $1.5$ & $0.113\pm0.002$ & $(2.064\pm0.196)\times10^0$ & $0.08/5$ & $0.134\pm0.001$ & $(7.117\pm0.375)\times10^0$   & $0.4/9$\\
      & $1.7$ & $0.094\pm0.002$ & $(1.293\pm0.169)\times10^0$ & $2/4$    & $0.133\pm0.001$ & $(6.053\pm0.354)\times10^0$   & $0.3/8$\\
\hline
      & $0.1$ & $0.137\pm0.002$ & $(9.281\pm0.516)\times10^0$ & $0.4/8$  & $0.128\pm0.001$ & $(1.480\pm0.088)\times10^{1}$ & $2/8$\\
      & $0.3$ & $0.137\pm0.002$ & $(9.562\pm0.601)\times10^0$ & $0.05/7$ & $0.129\pm0.002$ & $(1.512\pm0.090)\times10^{1}$ & $1/8$\\
      & $0.5$ & $0.130\pm0.002$ & $(8.807\pm0.511)\times10^0$ & $0.3/8$  & $0.133\pm0.002$ & $(1.566\pm0.092)\times10^{1}$ & $3/8$\\
      & $0.7$ & $0.135\pm0.002$ & $(8.375\pm0.527)\times10^0$ & $0.1/7$  & $0.149\pm0.002$ & $(1.543\pm0.086)\times10^{1}$ & $1/8$\\
      & $0.9$ & $0.136\pm0.001$ & $(7.181\pm0.381)\times10^0$ & $1/9$    & $0.143\pm0.002$ & $(1.331\pm0.075)\times10^{1}$ & $0.2/8$\\
17.3  & $1.1$ & $0.121\pm0.001$ & $(6.036\pm0.346)\times10^0$ & $1/44$   & $0.139\pm0.002$ & $(1.149\pm0.065)\times10^{1}$ & $0.4/8$\\
      & $1.3$ & $0.115\pm0.001$ & $(4.485\pm0.283)\times10^0$ & $2/8$    & $0.140\pm0.002$ & $(9.986\pm0.633)\times10^0$   & $0.1/7$\\
      & $1.5$ & $0.115\pm0.001$ & $(3.557\pm0.225)\times10^0$ & $0.5/8$  & $0.133\pm0.002$ & $(8.618\pm0.554)\times10^0$   & $1/7$\\
      & $1.7$ & $0.102\pm0.001$ & $(2.318\pm0.189)\times10^0$ & $4/6$    & $0.127\pm0.002$ & $(6.873\pm0.449)\times10^0$   & $0.4/7$\\
      & $1.9$ & $0.098\pm0.002$ & $(1.667\pm0.167)\times10^0$ & $0.1/5$  & $0.124\pm0.002$ & $(5.261\pm0.401)\times10^0$   & $1/6$\\
      & $2.1$ & $0.076\pm0.001$ & $(0.833\pm0.115)\times10^0$ & $2/4$    & $0.118\pm0.002$ & $(2.968\pm0.205)\times10^0$   & $6/7$\\
\hline
\end{tabular}%
\end{center}} \vskip1.0cm
\end{table*}

\begin{table*}[htbp] \vspace{0.cm} \justifying\noindent {\small Table 3. Same
as Table 1, but corresponding to the curves in Figures 5 and 6 for
$\bar p$ and $p$ respectively.\\} \vspace{-.5cm} {\small
\hspace{0.cm}
\begin{center}
\newcommand{\tabincell}[2]{\begin{tabular}{@{}#1@{}}#2\end{tabular}}
\begin{tabular} {cc|ccc|ccc}\\ \hline
$\sqrt{s}$ (GeV) & $y$ & \tabincell{c}{\\$T$ (GeV)} & \tabincell{c}{$\bar p$\\$V$ (fm$^3$)} &\tabincell{c}{\\$\chi^2$/ndof} & \tabincell{c}{\\$T$ (GeV)} & \tabincell{c}{$p$\\$V$ (fm$^3$)} & \tabincell{c}{\\$\chi^2$/ndof} \\
\hline
      & $0.1$ & $-$             & $-$                            & $-$    & $0.104\pm0.001$ & $(1.662\pm0.245)\times10^{1}$ & $4/7$\\
      & $0.3$ & $-$             & $-$                            & $-$    & $0.095\pm0.001$ & $(1.993\pm0.244)\times10^{1}$ & $9/10$\\
      & $0.5$ & $-$             & $-$                            & $-$    & $0.091\pm0.001$ & $(2.176\pm0.286)\times10^{1}$ & $8/10$\\
      & $0.7$ & $-$             & $-$                            & $-$    & $0.097\pm0.001$ & $(2.359\pm0.273)\times10^{1}$ & $17/10$\\
6.3   & $0.9$ & $-$             & $-$                            & $-$    & $0.089\pm0.001$ & $(3.633\pm0.520)\times10^{1}$ & $4/9$\\
      & $1.1$ & $-$             & $-$                            & $-$    & $0.090\pm0.001$ & $(3.326\pm0.630)\times10^{1}$ & $3/6$\\
      & $1.3$ & $-$             & $-$                            & $-$    & $0.074\pm0.001$ & $(4.321\pm0.963)\times10^{1}$ & $1/6$\\
      & $1.5$ & $-$             & $-$                            & $-$    & $0.064\pm0.001$ & $(4.180\pm1.066)\times10^{1}$ & $2/6$\\
      & $1.7$ & $-$             & $-$                            & $-$    & $0.043\pm0.001$ & $(2.855\pm1.225)\times10^{1}$ & $0.3/6$\\
\hline
      & $0.1$ & $0.114\pm0.002$ & $(3.326\pm0.458)\times10^{-1}$ & $9/7$  & $0.135\pm0.002$ & $(1.270\pm0.121)\times10^{1}$ & $1/9$\\
      & $0.3$ & $0.109\pm0.002$ & $(3.236\pm0.457)\times10^{-1}$ & $14/7$ & $0.133\pm0.002$ & $(1.405\pm0.136)\times10^{1}$ & $1/9$\\
      & $0.5$ & $0.109\pm0.002$ & $(2.025\pm0.332)\times10^{-1}$ & $10/6$ & $0.133\pm0.002$ & $(1.704\pm0.165)\times10^{1}$ & $1/9$\\
      & $0.7$ & $0.111\pm0.003$ & $(1.433\pm0.301)\times10^{-1}$ & $10/5$ & $0.125\pm0.002$ & $(2.061\pm0.211)\times10^{1}$ & $0.3/9$\\
7.7   & $0.9$ & $0.065\pm0.001$ & $(6.943\pm1.945)\times10^{-2}$ & $12/5$ & $0.124\pm0.002$ & $(2.637\pm0.272)\times10^{1}$ & $0.2/9$\\
      & $1.1$ & $-$             & $-$                            & $-$    & $0.126\pm0.002$ & $(3.075\pm0.302)\times10^{1}$ & $2/9$\\
      & $1.3$ & $-$             & $-$                            & $-$    & $0.114\pm0.001$ & $(3.436\pm0.384)\times10^{1}$ & $1/9$\\
      & $1.5$ & $-$             & $-$                            & $-$    & $0.104\pm0.001$ & $(3.801\pm0.505)\times10^{1}$ & $1/8$\\
      & $1.7$ & $-$             & $-$                            & $-$    & $0.086\pm0.001$ & $(4.103\pm0.783)\times10^{1}$ & $1/6$\\
\hline
      & $0.1$ & $0.105\pm0.003$ & $(3.835\pm1.138)\times10^{-1}$ & $6/3$  & $0.135\pm0.002$ & $(1.318\pm0.125)\times10^{1}$ & $0.2/9$\\
      & $0.3$ & $0.112\pm0.002$ & $(3.713\pm0.535)\times10^{-1}$ & $4/7$  & $0.137\pm0.002$ & $(1.417\pm0.134)\times10^{1}$ & $0.3/9$\\
      & $0.5$ & $0.104\pm0.002$ & $(2.982\pm0.647)\times10^{-1}$ & $4/5$  & $0.133\pm0.002$ & $(1.632\pm0.156)\times10^{1}$ & $1/9$\\
      & $0.7$ & $0.131\pm0.002$ & $(1.682\pm0.241)\times10^{-1}$ & $29/6$ & $0.127\pm0.002$ & $(1.928\pm0.194)\times10^{1}$ & $0.1/9$\\
8.8   & $0.9$ & $0.094\pm0.002$ & $(8.812\pm1.583)\times10^{-2}$ & $11/6$ & $0.128\pm0.002$ & $(2.374\pm0.236)\times10^{1}$ & $0.08/9$\\
      & $1.1$ & $-$             & $-$                          & $-$      & $0.127\pm0.002$ & $(2.931\pm0.293)\times10^{1}$ & $0.08/9$\\
      & $1.3$ & $-$             & $-$                          & $-$      & $0.125\pm0.002$ & $(3.365\pm0.342)\times10^{1}$ & $1/9$\\
      & $1.5$ & $-$             & $-$                          & $-$      & $0.112\pm0.001$ & $(3.630\pm0.411)\times10^{1}$ & $0.2/9$\\
      & $1.7$ & $-$             & $-$                          & $-$      & $0.097\pm0.001$ & $(3.748\pm0.584)\times10^{1}$ & $1/7$\\
      & $1.9$ & $-$             & $-$                          & $-$      & $0.085\pm0.001$ & $(3.785\pm0.730)\times10^{1}$ & $1/6$\\
\hline
      & $0.1$ & $0.100\pm0.001$ & $(9.128\pm1.170)\times10^{-1}$ & $11/9$ & $0.145\pm0.002$ & $(7.934\pm0.667)\times10^0$   & $1/10$\\
      & $0.3$ & $0.101\pm0.001$ & $(9.005\pm1.106)\times10^{-1}$ & $5/9$  & $0.148\pm0.002$ & $(8.276\pm0.678)\times10^0$   & $1/10$\\
      & $0.5$ & $0.104\pm0.002$ & $(7.481\pm0.905)\times10^{-1}$ & $2/9$  & $0.141\pm0.002$ & $(9.202\pm0.872)\times10^0$   & $1/9$\\
      & $0.7$ & $0.112\pm0.001$ & $(6.565\pm0.747)\times10^{-1}$ & $2/9$  & $0.145\pm0.002$ & $(1.100\pm0.101)\times10^{1}$ & $0.2/9$\\
12.3  & $0.9$ & $0.103\pm0.001$ & $(4.409\pm0.586)\times10^{-1}$ & $5/8$  & $0.142\pm0.002$ & $(1.324\pm0.105)\times10^{1}$ & $1/11$\\
      & $1.1$ & $0.111\pm0.001$ & $(2.464\pm0.303)\times10^{-1}$ & $4/8$  & $0.135\pm0.001$ & $(1.683\pm0.139)\times10^{1}$ & $0.1/11$\\
      & $1.3$ & $0.103\pm0.002$ & $(1.558\pm0.232)\times10^{-1}$ & $8/7$  & $0.132\pm0.001$ & $(2.091\pm0.177)\times10^{1}$ & $0.3/11$\\
      & $1.5$ & $0.100\pm0.002$ & $(7.733\pm1.344)\times10^{-2}$ & $11/6$ & $0.127\pm0.001$ & $(2.468\pm0.218)\times10^{1}$ & $1/11$\\
      & $1.7$ & $-$             & $-$                            & $-$    & $0.119\pm0.001$ & $(2.891\pm0.274)\times10^{1}$ & $2/11$\\
      & $1.9$ & $-$             & $-$                            & $-$    & $0.104\pm0.001$ & $(3.098\pm0.434)\times10^{1}$ & $1/8$\\
\hline
      & $0.1$ & $0.126\pm0.002$ & $(1.719\pm0.206)\times10^0$ & $0.4/8$ & $0.150\pm0.002$ & $(7.374\pm0.755)\times10^0$   & $0.2/8$\\
      & $0.3$ & $0.102\pm0.001$ & $(1.499\pm0.174)\times10^0$ & $11/10$ & $0.150\pm0.002$ & $(7.430\pm0.760)\times10^0$   & $0.2/8$\\
      & $0.5$ & $0.126\pm0.001$ & $(1.544\pm0.149)\times10^0$ & $3/10$  & $0.150\pm0.002$ & $(7.882\pm0.810)\times10^0$   & $0.08/8$\\
      & $0.7$ & $0.133\pm0.002$ & $(1.438\pm0.144)\times10^0$ & $1/9$   & $0.143\pm0.002$ & $(9.096\pm0.964)\times10^0$   & $0.2/8$\\
      & $0.9$ & $0.121\pm0.002$ & $(1.086\pm0.136)\times10^0$ & $2/8$   & $0.151\pm0.002$ & $(1.061\pm0.109)\times10^{1}$ & $0.5/8$\\
17.3  & $1.1$ & $0.127\pm0.002$ & $(0.805\pm0.111)\times10^0$ & $1/7$   & $0.141\pm0.002$ & $(1.191\pm0.149)\times10^{1}$ & $0.09/7$\\
      & $1.3$ & $0.123\pm0.002$ & $(0.568\pm0.066)\times10^0$ & $4/8$   & $0.138\pm0.002$ & $(1.555\pm0.196)\times10^{1}$ & $0.04/7$\\
      & $1.5$ & $0.161\pm0.005$ & $(0.410\pm0.081)\times10^0$ & $1/5$   & $0.135\pm0.002$ & $(1.950\pm0.248)\times10^{1}$ & $0.02/7$\\
      & $1.7$ & $0.136\pm0.005$ & $(0.214\pm0.064)\times10^0$ & $2/4$   & $0.136\pm0.002$ & $(2.515\pm0.392)\times10^{1}$ & $0.2/6$\\
      & $1.9$ & $0.071\pm0.003$ & $(0.057\pm0.029)\times10^0$ & $1/3$   & $0.142\pm0.003$ & $(2.868\pm0.491)\times10^{1}$ & $0.001/5$\\
      & $2.1$ & $-$             & $-$                         & $-$     & $0.132\pm0.003$ & $(3.033\pm0.531)\times10^{1}$ & $0.4/5$\\
\hline
\end{tabular}%
\end{center}} \vskip1.0cm
\end{table*}

\end{document}